\documentclass[12pt]{article}

\usepackage[margin=3cm]{geometry}
\usepackage{tikz}
\usetikzlibrary{backgrounds}
\usepackage{xspace}
\usepackage{tikz-cd}
\RequirePackage{subcaption}
\usepackage{standalone}
\usepackage[format=plain,
            labelfont={bf},
            textfont=it]{caption}
\usepackage{amsmath}
\usepackage{amsthm}
\usepackage{amssymb}
\usepackage{etoolbox}
\usepackage{hyperref}
\usepackage{caption}
\usepackage[numbers, sort]{natbib}

\newcommand{\id}{\ensuremath{\operatorname{id}}}
\newcommand{\ord}[1]{\underline{#1}}
\newcommand{\op}{\mathsf{op}}

\newcommand{\Set}{\text{Set}}
\newcommand{\PSh}[1]{\lbrack {#1}^\op, \Set \rbrack}

\newcommand\hio{\emph{homotopy.io}\xspace}
\def\N{\ensuremath{\mathbb{N}}\xspace}
\def\vpgh{\vphantom{gh}}

\newcommand\cat[1]{\ensuremath{\mathbf{#1}}\xspace}
\def\bu{{\color{gray}\text{\tiny\ensuremath{\bullet}}}}

\newcommand{\Deg}{\operatorname{Deg}}
\newcommand{\Sub}{\operatorname{Sub}}
\newcommand{\reg}[1]{\mathsf{r}\hspace{1pt}#1}
\newcommand{\sing}[1]{\mathsf{s}\hspace{1pt}#1}
\newcommand{\singmap}[1]{{#1}^\mathsf{s}}
\newcommand{\regmap}[1]{{#1}^\mathsf{r}}

\tikzset{%
  cell/.style = {draw=black,fill=white,rectangle},
  height/.style = {densely dotted, draw=black, fill=none, shorten <=0.05cm, shorten >=0.05cm},
  rewrite/.style = {->, draw=black, shorten <=0.2cm, shorten >=0.2cm}
}

\newcounter{commcounter}
\theoremstyle{theorem} 
\newtheorem{theorem}{Theorem}
\newtheorem{lemma}[theorem]{Lemma}
\newtheorem{proposition}[theorem]{Proposition}

\theoremstyle{definition}

\newtheorem{definition}[theorem]{Definition}
\newtheorem{example}[theorem]{Example}

\def\xo{.3}
\def\yo{.3}
\tikzset{v/.style={fill=white, circle, draw=none, inner sep=0pt}}
\tikzset{e/.style={double, white, double=black!50, line width=1pt, line width=1.0pt}}
\tikzset{f/.style={line width=.7pt, black!50}}

\newcommand\ee[2]{\draw [e] (#1) to (#2);}
\newcommand\ff[2]{\draw [f] (#1) to (#2);}

\usepackage{enumitem}
\setlist[itemize]{leftmargin=23pt}
\setlist[enumerate]{leftmargin=23pt}

\def\suckpara{\vspace{0pt}}
\def\sucksec{\vspace{0pt}}

\makeatletter
\def\calign@preamble{%
   &\hfil\strut@
    \setboxz@h{\@lign$\m@th\displaystyle{##}$}%
    \ifmeasuring@\savefieldlength@\fi
    \set@field
    \hfil
    \tabskip\alignsep@
}
\let\cmeasure@\measure@
\patchcmd\cmeasure@{\divide\@tempcntb\tw@}{}{}{}
\patchcmd\cmeasure@{\divide\@tempcntb\tw@}{}{}{}
\patchcmd\cmeasure@{\ifodd\maxfields@
  \global\advance\maxfields@\@ne
  \fi}{}{}{}

\makeatother

\newcommand\ignore[1]{}

\begin{document}

\title{\bf Zigzag normalisation for\\associative $n$-categories}

\author{Lukas Heidemann\footnote{University of Oxford, \texttt{lukas.heidemann@cs.ox.ac.uk}}~, David Reutter\footnote{University of Hamburg, \texttt{david.reutter@uni-hamburg.de}} ~and Jamie Vicary\footnote{University of Cambridge, \texttt{jamie.vicary@cl.cam.ac.uk}}}

\date{May 10, 2022}

\maketitle

\begin{abstract}
The theory of associative $n$-categories has recently been proposed as a strictly associative and unital approach to higher category theory. As a foundation for a proof assistant, this is potentially attractive, since it has the potential to allow simple formal proofs of complex high-dimensional algebraic phenomena. However, the  theory relies on an implicit  term normalisation procedure to recognize correct composites, with no recursive method available for computing~it.

Here we describe a new approach to term normalisation in associative $n$\-categories, based on the categorical zigzag construction. This radically simplifies the theory, and yields a recursive algorithm for normalisation, which we prove is correct. Our use of categorical lifting properties allows us to give efficient proofs of our results. This normalisation algorithm forms a core component of the proof assistant \hio, and we illustrate our scheme with worked examples.
\end{abstract}

\section{Introduction}

\subsection{Overview}

\paragraph{Motivation}
The flexibility of weak higher categories has enabled their wide use across  many areas of mathematics, computer science, and physics. The most well-known include  the homotopy type theory programme on univalent foundations for mathematics~\cite{hottbook, Awodey_2008, Voevodsky_2006}, motivated by the intensional groupoid model for Martin-L\"of type theory~\cite{Hofmann_1994}; Lurie's outline proof~\cite{Lurie_2008} of the cobordism hypothesis of Baez and Dolan~\cite{Baez_1995}, and the associated new perspective it brought for topological quantum field theory~\cite{Atiyah_1990, CSPthesis}; and the higher topos theory programme~\cite{Lurie_2009}, with broad implications for both logic and geometry, which develops ideas going back to Grothendieck~\cite{Grothendieck_1983}. In computer science, other applications include rewriting~\cite{Mimram_2014, Lafont_1997, Guiraud_2012}, quantum computation~\cite{Reutter_2016, Jaffe_2016}, and concurrency~\cite{Bruni_2002, Goubault_2003}.

In a weak higher category, equations hold only up to higher coherence data, which itself satisfies further equations up to coherence data, and so on ad infinitum, yielding a bureaucratic syntax in which conceptually simple proofs can become long-winded. {Traditionally, this has been the price that must be paid for proof-relevance. {Strict} models~\cite[Section 1.4]{Leinster_2004} discard this coherence data, but at the cost of expressivity, since not every weak higher category is equivalent to  a strict one.

\textit{Associative $n$-categories} (ANCs) are a new \textit{semistrict} model that aims to strike a balance between these extremes: having enough strictness for practical use, while retaining sufficient weakness to remain conjecturally equivalent to the fully general case~\cite{Dorn_2018, Dorn_2022, Reutter_2019}. However, the original theory of ANCs cannot be directly implemented, in particular lacking an algorithm for \textit{term normalisation}, a key part of the theory which allows recognition of valid composites.

A new theory of normalisation is therefore required, one which is well-adapted to the data structures of a potential proof assistant, and with respect to which an recursive algorithm for normalisation can be provided. We develop this new theory here, and describe its role within an implementation, with the goal of making higher category theory more accessible for the working computer scientist.

\begin{figure*}
\tikzset{dot/.style={draw=black, circle, fill=white, inner sep=0pt, font=\tiny, minimum width=10pt, fill=white, text=black}}
\tikzset{bluedot/.style={dot, fill=blue!50, text=black, fill=white, text=black}}
\tikzset{yellowdot/.style={dot, fill=yellow!50, text=black}}
\tikzset{bluedot/.style={dot, fill=blue!50, text=black}}
\tikzset{2cell/.style={draw=black, circle, fill=red!50, inner sep=0pt, minimum width=10pt, fill=white, text=black, thick}}
\tikzset{scaffold/.style={->, shorten <=3pt, shorten >=3pt, black, thick}}
\tikzset{equality/.style={shorten <=5pt, shorten >=5pt, double distance=2pt}}
\tikzset{string/.style={black, thick}}
\tikzset{vertical/.style={}}
\tikzset{regular/.style={opacity=0}}
\tikzset{equality/.style={opacity=0}}
\tikzset{horizontal/.style={opacity=0}}
\newcommand\only[1]{}
$$
\begin{aligned}
\begin{tikzpicture}[scale=0.5,yscale=1.2, xscale=1, yscale=1.2, scale=1]
\path [red, ultra thick, use as bounding box] (0,0) rectangle +(10,6);
\tikzset{dot/.style={}}
\node [bluedot, regular] (00) at (0,0) {};
\node [dot] (10) at (1,0) {};
\node [dot, regular] (20) at (2.5,0) {};
\node [dot] (30) at (4,0) {};
\node [dot, regular] (40) at (6.5,0) {};
\node [dot] (50) at (9,0) {};
\node [dot, regular] (60) at (10,0) {};
\node [bluedot, regular] (01) at (0,1) {};
\node [dot] (11) at (1,1) {};
\node [dot, regular] (21) at (2.5,1) {};
\node [2cell] (31) at (4,1) {};
\node [dot, regular] (41) at (6.5,1) {};
\node [2cell] (51) at (9,1) {};
\node [dot, regular] (61) at (10,1) {};
\node [bluedot, regular] (02) at (0,2) {};
\node [dot] (12) at (1,2) {};
\node [dot, regular] (22) at (2,2) {};
\node [dot] (32) at (3,2) {};
\node [dot, regular] (42) at (4,2) {};
\node [dot] (52) at (5,2) {};
\node [dot, regular] (62) at (7,2) {};
\node [dot] (72) at (9,2) {};
\node [dot, regular] (82) at (10,2) {};
\node [bluedot, regular] (03) at (0,3) {};
\node [2cell] (13) at (2,3) {};
\node [dot, regular] (23) at (3.5,3) {};
\node [dot] (33) at (5,3) {};
\node [dot, regular] (43) at (7,3) {};
\node [dot] (53) at (9,3) {};
\node [dot, regular] (63) at (10,3) {};
\node [bluedot, regular] (04) at (0,4) {};
\node [dot] (14) at (2,4) {};
\node [dot, regular] (24) at (3.5,4) {};
\node [dot] (34) at (5,4) {};
\node [dot, regular] (44) at (7,4) {};
\node [dot] (54) at (9,4) {};
\node [dot, regular] (64) at (10,4) {};
\node [bluedot, regular] (05) at (0,5) {};
\node [dot] (15) at (2,5) {};
\node [dot, regular] (25) at (3.5,5) {};
\node [dot] (35) at (5,5) {};
\node [dot, regular] (45) at (6,5) {};
\node [2cell] (55) at (7,5) {};
\node [dot, regular] (65) at (8,5) {};
\node [dot] (75) at (9,5) {};
\node [dot, regular] (85) at (10,5) {};
\node [bluedot, regular] (06) at (0,6) {};
\node (16) at (2,6) {};
\node (36) at (5,6) {};
\node (56) at (9,6) {};
\node (66) at (10,6) {};
\draw [string] (30.center) to (31.center);
\draw [string] (31.center) to [out=left, in=down] (32.center) to [out=up, in=right] (13.center);
\draw [string] (10.center) to (12.center) to [out=up, in=left] (13.center) to (14.center) to (15.center) to (16.center);
\draw [string] (31.center) to [out=right, in=down] (52.center) to (33.center) to (34.center) to (35.center) to (36.center);
\draw [string] (50.center) to (51.center) to (72.center) to (53.center) to (54.center) to (75.center) to (56.center);
\node [2cell] (55) at (7,5) {};
\node [2cell] (31) at (4,1) {};
\node [2cell] (13) at (2,3) {};
\node [2cell] (51) at (9,1) {};
\begin{pgfonlayer}{background}
\draw [fill=black!5, draw=none] (00.center) rectangle (66.center);
\end{pgfonlayer}
\end{tikzpicture}
\end{aligned}
\hspace{2cm}
\begin{aligned}
\begin{tikzpicture}[scale=0.5,yscale=1.2, xscale=1, yscale=1.2, scale=1]
\path [red, ultra thick, use as bounding box] (0,0) rectangle +(10,6);
\tikzset{regular/.style={opacity=1}}
\tikzset{vertical/.style={opacity=1}}
\tikzset{dot/.style={opacity=1, inner sep=0pt, font=\scriptsize\bf}}
\tikzset{scaffold/.style={->, shorten <=1.5pt, shorten >=1.5pt, black, thick}}
\tikzset{horizontal/.style={opacity=1}}
\tikzset{equality/.style={scaffold}}
\node [dot, regular] (00) at (0,0) {0};
\node [dot] (10) at (1,0) {1};
\node [dot, regular] (20) at (2.5,0) {0};
\node [dot] (30) at (4,0) {1};
\node [dot, regular] (40) at (6.5,0) {0};
\node [dot] (50) at (9,0) {1};
\node [dot, regular] (60) at (10,0) {0};
\node [dot, regular] (01) at (0,1) {0};
\node [dot] (11) at (1,1) {1};
\node [dot, regular] (21) at (2.5,1) {0};
\node [dot] (31) at (4,1) {2};
\node [dot, regular] (41) at (6.5,1) {0};
\node [dot] (51) at (9,1) {2};
\node [dot, regular] (61) at (10,1) {0};
\node [dot, regular] (02) at (0,2) {0};
\node [dot] (12) at (1,2) {1};
\node [dot, regular] (22) at (2,2) {0};
\node [dot] (32) at (3,2) {1};
\node [dot, regular] (42) at (4,2) {0};
\node [dot] (52) at (5,2) {1};
\node [dot, regular] (62) at (7,2) {0};
\node [dot] (72) at (9,2) {1};
\node [dot, regular] (82) at (10,2) {0};
\node [dot, regular] (03) at (0,3) {0};
\node [dot] (13) at (2,3) {2};
\node [dot, regular] (23) at (3.5,3) {0};
\node [dot] (33) at (5,3) {1};
\node [dot, regular] (43) at (7,3) {0};
\node [dot] (53) at (9,3) {1};
\node [dot, regular] (63) at (10,3) {0};
\node [dot, regular] (04) at (0,4) {0};
\node [dot] (14) at (2,4) {1};
\node [dot, regular] (24) at (3.5,4) {0};
\node [dot] (34) at (5,4) {1};
\node [dot, regular] (44) at (7,4) {0};
\node [dot] (54) at (9,4) {1};
\node [dot, regular] (64) at (10,4) {0};
\node [dot, regular] (05) at (0,5) {0};
\node [dot] (15) at (2,5) {1};
\node [dot, regular] (25) at (3.5,5) {0};
\node [dot] (35) at (5,5) {1};
\node [dot, regular] (45) at (6,5) {0};
\node [dot] (55) at (7,5) {2};
\node [dot, regular] (65) at (8,5) {0};
\node [dot] (75) at (9,5) {1};
\node [dot, regular] (85) at (10,5) {0};
\node [dot, regular] (06) at (0,6) {0};
\node [dot] (16) at (2,6) {1};
\node [dot, regular] (26) at (3.5,6) {0};
\node [dot] (36) at (5,6) {1};
\node [dot, regular] (46) at (7,6) {0};
\node [dot] (56) at (9,6) {1};
\node [dot, regular] (66) at (10,6) {0};
\draw [scaffold, horizontal] (00) to (10);
\draw [scaffold, horizontal] (20) to (10);
\draw [scaffold, horizontal] (20) to (30);
\draw [scaffold, horizontal] (40) to (50);
\draw [scaffold, horizontal] (40) to (30);
\draw [scaffold, horizontal] (60) to (50);
\draw [scaffold, horizontal] (01) to (11);
\draw [scaffold, horizontal] (21) to (11);
\draw [scaffold, horizontal] (21) to (31);
\draw [scaffold, horizontal] (41) to (31);
\draw [scaffold, horizontal] (41) to (51);
\draw [scaffold, horizontal] (61) to (51);
\draw [scaffold, horizontal] (02) to (12);
\draw [scaffold, horizontal] (22) to (12);
\draw [scaffold, horizontal] (22) to (32);
\draw [scaffold, horizontal] (42) to (32);
\draw [scaffold, horizontal] (42) to (52);
\draw [scaffold, horizontal] (62) to (52);
\draw [scaffold, horizontal] (62) to (72);
\draw [scaffold, horizontal] (82) to (72);
\draw [scaffold, horizontal] (03) to (13);
\draw [scaffold, horizontal] (23) to (13);
\draw [scaffold, horizontal] (23) to (33);
\draw [scaffold, horizontal] (43) to (33);
\draw [scaffold, horizontal] (43) to (53);
\draw [scaffold, horizontal] (63) to (53);
\draw [scaffold, horizontal] (04) to (14);
\draw [scaffold, horizontal] (24) to (14);
\draw [scaffold, horizontal] (24) to (34);
\draw [scaffold, horizontal] (44) to (34);
\draw [scaffold, horizontal] (44) to (54);
\draw [scaffold, horizontal] (64) to (54);
\draw [scaffold, horizontal] (05) to (15);
\draw [scaffold, horizontal] (25) to (15);
\draw [scaffold, horizontal] (25) to (35);
\draw [scaffold, horizontal] (45) to (35);
\draw [scaffold, horizontal] (45) to (55);
\draw [scaffold, horizontal] (65) to (55);
\draw [scaffold, horizontal] (65) to (75);
\draw [scaffold, horizontal] (85) to (75);
\draw [scaffold, horizontal] (06) to (16);
\draw [scaffold, horizontal] (26) to (16);
\draw [scaffold, horizontal] (26) to (36);
\draw [scaffold, horizontal] (46) to (36);
\draw [scaffold, horizontal] (46) to (56);
\draw [scaffold, horizontal] (66) to (56);
\draw [equality, vertical] (00) to (01);
\draw [scaffold, vertical] (10) to (11);
\draw [equality, vertical] (20) to (21);
\draw [scaffold, vertical] (30) to (31);
\draw [equality, vertical] (40) to (41);
\draw [scaffold, vertical] (50) to (51);
\draw [equality, vertical] (60) to (61);
\draw [equality, vertical] (02) to (01);
\draw [scaffold, vertical] (12) to (11);
\draw [equality, vertical] (22) to (21);
\draw [scaffold, vertical] (32) to (31);
\draw [scaffold, vertical] (52) to (31);
\draw [equality, vertical] (62) to (41);
\draw [scaffold, vertical] (72) to (51);
\draw [equality, vertical] (82) to (61);
\draw [equality, vertical] (02) to (03);
\draw [scaffold, vertical] (12) to (13);
\draw [scaffold, vertical] (32) to (13);
\draw [equality, vertical] (42) to (23);
\draw [scaffold, vertical] (52) to (33);
\draw [equality, vertical] (62) to (43);
\draw [scaffold, vertical] (72) to (53);
\draw [equality, vertical] (82) to (63);
\draw [equality, vertical] (04) to (03);
\draw [scaffold, vertical] (14) to (13);
\draw [equality, vertical] (24) to (23);
\draw [scaffold, vertical] (34) to (33);
\draw [equality, vertical] (44) to (43);
\draw [scaffold, vertical] (54) to (53);
\draw [equality, vertical] (64) to (63);
\draw [equality, vertical] (04) to (05);
\draw [scaffold, vertical] (14) to (15);
\draw [equality, vertical] (24) to (25);
\draw [scaffold, vertical] (34) to (35);
\draw [equality, vertical] (44) to (45);
\draw [equality, vertical] (44) to (65);
\draw [scaffold, vertical] (54) to (75);
\draw [equality, vertical] (64) to (85);
\draw [equality, vertical] (06) to (05);
\draw [scaffold, vertical] (16) to (15);
\draw [equality, vertical] (26) to (25);
\draw [scaffold, vertical] (36) to (35);
\draw [equality, vertical] (46) to (45);
\draw [equality, vertical] (46) to (65);
\draw [scaffold, vertical] (56) to (75);
\draw [equality, vertical] (66) to (85);
\end{tikzpicture}
\end{aligned}
$$

\vspace{-5pt}
\caption{Representing a 2-dimensional string diagram as an  iterated zigzag of natural numbers.\label{fig:zigzagmotivation}}
\end{figure*}
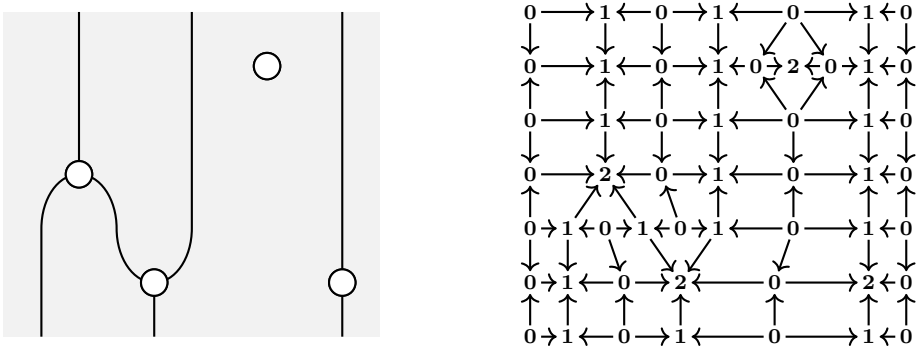

\suckpara
\paragraph{Diagrams.} Associative $n$-categories have the striking feature of being inherently geometrical, with terms in the theory having a direct geometrical representation. Every term has a dimension, and the terms of dimension $n$ are called \textit{$n$\-diagrams}. A 0-diagram is a point, a 1-diagram is a sequence of points arranged on a line, and a 2-diagram is a combinatorial version of a planar string diagram~\cite{selingersurvey}. In the general case, an $n$\-diagram can be interpreted as a combinatorial ``$n$-dimensional string diagram''.

At LICS~2019 a simple inductive term model for these $n$\-diagrams was presented~\cite{Reutter_2019}, called \emph{zigzags}, which we make further use of here. An example of a 2\-diagram is shown on the left in Figure~\ref{fig:zigzagmotivation}, with its underlying zigzag structure shown on the right, the natural numbers giving the dimension of the component at each point. The zigzag representation makes the combinatorial structure explicit, but we will generally prefer the cleaner visual style of the left image.

\suckpara
\paragraph{Normalisation.}
Associative $n$\-categories are strictly associative and strictly unital in all dimensions, two attractive properties which remove considerable bureaucracy from proof construction. However, the theory gains these properties in very different ways. The strict associativity is explicit: given composable 1\-morphisms $f,g,h$, the composites\footnote{Here and throughout we use forward composition notation.} $(f \cdot g) \cdot h$ and $f \cdot (g \cdot h)$  are syntactically identical, and can be drawn as the following 1-diagram:
\tikzset{blob/.style={draw, circle, fill, inner sep=0pt, minimum width=4pt}}
\begin{equation}
\begin{tikzpicture}
\draw [thick] (0.25,0) to (3.75,0);
\node [blob] at (1,0) {};
\node [blob] at (2,0) {};
\node [blob] at (3,0) {};
\node at (1,.4) {$\vphantom{gh}f$};
\node at (2,.4) {$\vpgh g$};
\node at (3,.4) {$h$};
\end{tikzpicture}
\end{equation}
In contrast, the composite $f \cdot \id$ is not syntactically identical to $f$. Instead, there is a nontrivial \textit{diagram normalisation}
process $f \cdot \id \leadsto f$:
\begin{equation}
\label{eq:norm1d}
\begin{tikzpicture}
\draw [thick] (0.25,0) to (2.75,0);
\node [blob] at (1,0) {};
\node [blob, red] at (2,0) {};
\node at (1,.4) {$\vphantom{gh}f$};
\node [red] at (2,.4) {$\id$};
\end{tikzpicture}
\qquad\leadsto\qquad
\begin{tikzpicture}
\draw [thick] (0.25,0) to (1.75,0);
\node [blob] at (1,0) {};
\node at (1,.4) {$\vphantom{gh}f$};
\end{tikzpicture}
\end{equation}
This normalisation process removes identity structures, yielding a ``strictly unital'' form for the composite, in this case $f$ itself. To give the user the experience of interacting with a strictly unital theory, the proof assistant performs normalisation silently after every user interaction, ensuring the user sees only the normal form.

In low dimensions, normalisation seems to be a simple process. In dimension 1, a composite is given by a string of tokens, and normalisation simply removes any identity tokens, as shown above in expression~\eqref{eq:norm1d}. In dimension 2, a composite can be understood as a string diagram in a weakly unital monoidal category, where some strands are explicitly labelled by the unit object, indicated here by dotted red wires. In this case the normalisation process removes these unit structures:
$$
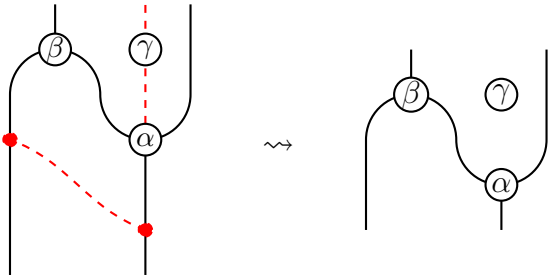
\begin{aligned}
\begin{tikzpicture}[thick, scale=.6]
\tikzset{blob/.style={draw, circle, fill, inner sep=0pt, minimum width=5pt, font=\small}}
\draw (0,1) to [out=right, in=down] (1,2) to (1,4);
\draw (-2,3) to [out=left, in=up] (-3,2) to (-3,-2);
\draw [red, dashed] (0,1) to [] (0,3) node [black, blob, fill=white, solid, thick, minimum width=12pt] {$\gamma$} to (0,4);
\draw (0,-2) to (0,1) node [blob, fill=white, solid, minimum width=12pt] {$\alpha$} to [out=left, in=down] (-1.0,2) to [out=up, in=right] (-2,3) node [blob, fill=white, minimum width=12pt] {$\beta$} to (-2,4);
\draw [red, dashed] (0,-1) node [blob] {} to [out=160, in=-20] (-3,1) node [blob] {};
\end{tikzpicture}
\end{aligned}
\qquad
\leadsto
\qquad
\begin{aligned}
\begin{tikzpicture}[thick, scale=.6]
\tikzset{blob/.style={draw, circle, fill, inner sep=0pt, minimum width=5pt}}
\draw (-2,3) to [out=left, in=up] (-3,2) to (-3,0);
\draw (0,1) to [out=right, in=down] (1,2) to (1,4);
\draw (0,0) to (0,1) node [blob, fill=white, solid, minimum width=12pt] {$\alpha$} to [out=left, in=down] (-1,2) to [out=up, in=right] (-2,3) node [blob, fill=white, solid, minimum width=12pt] {$\beta$} to (-2,4);
\node [blob, fill=white, solid, minimum width=12pt] at (0,3) {$\gamma$};
\end{tikzpicture}
\end{aligned}
$$
The situation is similar in dimension 3; once again, normalisation simply removes all identity structures. These examples quickly  give us the impression that all identity structures are \emph{redundant}, since normalisation produces an algebraically simpler version of the diagram that omits them.

However, in dimension 4 and above, more subtle behaviour arises. The geometrical structure of the diagram can cause certain identity structures to become ``locked'', in a way that prevents them being removed by normalisation. If such an identity were removed, the resulting diagram would be algebraically ill-defined. We call these \textit{essential identities}, to contrast with the redundant identities we visualised above. We will see in Section~\ref{sec:computing} why these essential identities
 arise. This makes a normalisation algorithm in the general case non-obvious, since the naive strategy of ``removing all identity structures'' cannot succeed.

\begin{figure*}[t]
$$\begin{tikzpicture}[xscale=4, yscale=1.5]
\tikzset{topic/.style={draw, minimum width=3.4cm, thick, minimum height=1cm}}
\node (1) [topic] at (0,0) {\vphantom{qP}Term Formation};
\node at (0,-.6) {\em Contraction};
\node (2) [topic] at (1,0) {\vphantom{qP}Type Checking};
\node at (1,-.6) {\em Normalisation};
\node (3) [topic] at (2,0) {\vphantom{qP}Rendering};
\node at (2,-.6) {\em Geometrization};
\draw [thick, ->] (1.east) to (2.west);
\draw [thick, ->] (2.east) to (3.west);
\end{tikzpicture}$$

\vspace{-5pt}
\caption{An simplified overview of the homotopy.io processing pipeline.}
\label{fig:pipeline}
\end{figure*}
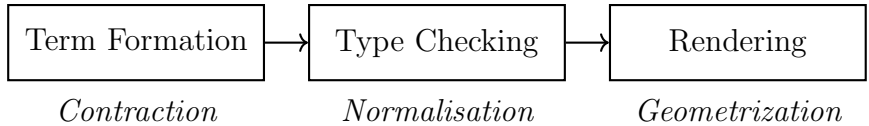

\suckpara
\paragraph{Type Checking.}
Given an $n$\-diagram, a question of central importance is whether it \textit{type-checks} with respect to some given signature $\Sigma$; that is, whether it correctly encodes an $n$\-morphism in the free ANC generated by $\Sigma$. The existing theory of ANCs gives a simple answer to this question: break the diagram into atomic ``pieces'', and for each piece, check that it normalizes to give an element of the signature. Normalisation therefore plays a critical role in this  aspect of the theory, and is essential for any implementation.

More details of the type-checking scheme are given in Section~\ref{sec:implementation}, where we also illustrate our normalisation algorithm in detail, using two substantial examples of real interest in higher category theory: the 3\-dimensional braiding, and the 5\-dimensional syllepsis.


\suckpara
\paragraph{Our\hspace{3pt}Contribution.}
We introduce a new mathematical foundation for term normalisation in associative $n$\-categories, focusing on the categorical properties of \textit{degeneracy maps} in categories of $n$\-diagrams. Defined in terms of a simple categorical lifting property, these degeneracy maps can be interpreted as  ``injecting identity structure'' into a $n$\-diagram; a  degeneracy map $f:D \to D'$ therefore serves as a witness that $D$ and $D'$ are similar, except $D$ contains fewer redundant identities. Proposition~\ref{lemma:degeneracy-pullback} shows that for any $n$\-diagram $D,$ and any pair of degeneracy maps $f:A \to D$ and $g:B \to D$, the pullback $A \times_D B$ exists, with the resulting map $A \times_D B \to D$  again a degeneracy map, which can be considered the ``joint resolution'' of $f,g$. In this way, the normalisation of $D$ can be characterized as the joint resolution of all degeneracy maps into $D$. Since there only finitely many up to isomorphism, this is well-defined.

We then show how normalisation can be computed. Given $n$\-diagrams $D$ and $A_1, \ldots, A_n$, equipped with $n$\-diagram maps $f_i:A_i \to D$, our central observation is that we can give a recursive algorithm for the \emph{relative normalisation} of $D$ with respect to the $f_i$, written $N$, as follows:
\begin{equation}
\nonumber
\begin{aligned}
\begin{tikzpicture}[xscale=.5, yscale=.7, scale=1]
\node (1) at (0,0) {$A_1$};
\node (2) at (1,1) {$A_2$};
\node [rotate=90]  at (2,2) {$\ddots$};
\node (n) at (3,3) {$A_n$};
\node (d) at (7,1.5) {$D$};
\draw [->] (1) to node [below right, font=\scriptsize] {$f_1$} (d.south west);
\draw [->] (2) to node [above left=-2pt, pos=.61, font=\scriptsize] {$f_2$} (d.west);
\draw [->] (n) to node [above right, pos=.3, font=\scriptsize] {$f_n$} (d.north west);
\end{tikzpicture}
\end{aligned}
\hspace{.2cm}\leadsto\hspace{-.2cm}
\begin{aligned}
\begin{tikzpicture}[xscale=.5, yscale=.7, scale=1]
\node (1) at (0,0) {$A_1$};
\node (2) at (1,1) {$A_2$};
\node [rotate=90]  at (2,2) {$\ddots$};
\node (n) at (3,3) {$A_n$};
\node (d) at (5,1.5) {$N$};
\node (d2) at (8,1.5) {$D$};
\draw [->] (1) to node [below right, pos=.44, font=\scriptsize] {$f_1'$} (d.south west);
\draw [->] (2) to node [above left=-2pt, pos=.6, font=\scriptsize] {$f_2'$} (d.west);
\draw [->] (n) to node [above right, pos=.3, font=\scriptsize] {$f_n'$} (d.north west);
\draw [->] (d) to node [above] {$d$} (d2);
\end{tikzpicture}
\end{aligned}
\end{equation}
Here $d$ is a degeneracy map, and we have $f_i' \cdot d = f_i$ for all \mbox{$i \in I$}, where $I = \{1, \ldots, n\}$.\footnote{We emphasize that the object $N$ and morphism $d:N \to D$ depend on the choice of morphisms $f_i$, although we suppress this in the notation.} Recalling the standard categorical definition of a \emph{sink} as an object equipped with a family of incoming morphisms,  the relative normalisation  provides a universal factorization of the sink $(D, \{ f_i : A_i \to D\}_{i \in I} )$  into  a composite of the sink $(N,\{ f_i': A_i \to N\}_{i \in I})$ with the morphism~$d$. 

We then compute the \textit{absolute normalisation} of a diagram $D$ as the relative normalisation of the sink $(D,\emptyset)$ consisting of $D$ equipped with the empty collection of morphisms.

\suckpara
\paragraph{Implementation.} 
We have implemented our results, and they form a central part of a proof assistant for associative $n$\-categories, called \hio. Implemented as a client-side web application, it is hosted at the following URL:
$$\text{\url{http://homotopy.io}}$$
The tool was launched in January 2019, and has since been loaded 12,000 times by over 4,000 users. It allows direct construction and manipulation of higher-categorical composites by a click-and-drag mechanic. In Section~\ref{sec:implementation} we provide links to proof objects, which illustrate some of our results.

We give a simplified overview of the proof assistant's processing pipeline in Figure~\ref{fig:pipeline}, which we  summarize as follows.

\vspace{3pt}
\noindent
-- \textit{Term Formation} is performed  primarily via the {contraction} mechanism, which allows part of an $n$\-diagram to be homotopically reduced. The theoretical foundation for this technique was presented at LICS 2019~\cite{Reutter_2019}.

\vspace{3pt}
\noindent
-- \textit{Type Checking} verifies that the term generated by the user interaction step is valid. The major component of this type checker is an implementation of  the recursive sink normalisation algorithm that we describe in this paper.

\vspace{3pt}
\noindent
-- \emph{Rendering} takes place via a geometrization process that extracts a cubical mesh from the term representation, which is then processed and sent to the video card for rendering. This component will be described in future work. 

\vspace{3pt}
\noindent
Since this is a theoretical article we will not present further details here of the implementation.

\sucksec\subsection{Related work}
\label{sec:relatedwork}

The theory of associative $n$\-categories was originally developed by Dorn, Douglas and Vicary~\cite{Dorn_2022}, and has been described in the thesis of Dorn~\cite{Dorn_2018} in terms of bundles of singular $n$\-cubes. We present a new approach that follows the zigzag construction of Reutter and Vicary~\cite{Reutter_2019}, giving us access to a simple inductive structure on terms. The theory of normalisation developed here makes heavy use of categorical lifting properties (cartesian and cocartesian maps),  categorical ``power tools'' which allow an efficient formal development. This also allows us to give a recursive algorithm for normalisation, which is not achieved in the singular $n$\-cubes approach.


While we believe our theory is in principle equivalent to that proposed by Dorn, Douglas and Vicary, we  make our constructions from first principles, giving a self-contained development. Our approach also has the advantage of allowing a concise presentation.

\sucksec
\subsection{Acknowledgements}

The authors are grateful to Eric Finster, Christoph Dorn and Christopher Douglas for useful discussions.

\sucksec
\subsection{Notation}

For any $n \geq 0$ we denote by $\ord{n}$ the finite total order \mbox{$\{ 0, \ldots, n - 1\}$}. We write $\Delta_+$ for the category where objects are these finite total orders, and morphisms are order-preserving maps. We also write $\Delta_=$ for the subcategory of
non-empty total orders, with maps that preserve the initial and final elements. For an order-preserving map $f: \ord n \to \ord m$ and some $i \in \ord m$, we write $f^{-1}(i)$ for its preimage as a subset of $\ord n$. The terminal category is denoted as $1$.

\sucksec
\section{The zigzag construction}
\label{sec:zigzags}

We begin by recalling the theory of categorical zigzags due to Reutter and Vicary~\cite{Reutter_2019}. Our presentation is in fact a mild generalization, permitting non-identity boundary maps, as we make clear below. The main object of study is the zigzag, defined as follows.

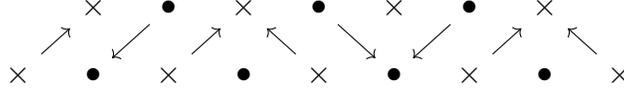
\begin{figure*}[t]
\begin{center}
\begin{tikzpicture}[yscale=.9]
  \node (r00) at (1, 1) {$\times$};
  \node (s00) at (2, 1) {$\bullet$};
  \node (r01) at (3, 1) {$\times$};
  \node (s01) at (4, 1) {$\bullet$};
  \node (r02) at (5, 1) {$\times$};
  \node (s02) at (6, 1) {$\bullet$};
  \node (r03) at (7, 1) {$\times$};

  \node (r10) at (0, 0) {$\times$};
  \node (s10) at (1, 0) {$\bullet$};
  \node (r11) at (2, 0) {$\times$};
  \node (s11) at (3, 0) {$\bullet$};
  \node (r12) at (4, 0) {$\times$};
  \node (s12) at (5, 0) {$\bullet$};
  \node (r13) at (6, 0) {$\times$};
  \node (s13) at (7, 0) {$\bullet$};
  \node (r14) at (8, 0) {$\times$};

  \draw[->] (s00) -- (s10);
  \draw[->] (s01) -- (s12);
  \draw[->] (s02) -- (s12);
  \draw[->] (r10) -- (r00);
  \draw[->] (r11) -- (r01);
  \draw[->] (r12) -- (r01);
  \draw[->] (r13) -- (r03);
  \draw[->] (r14) -- (r03);
\end{tikzpicture}
\end{center}

\vspace{-5pt}
  \caption{
    A monotone map $f : \ord{3} \to \ord{4}$ in $\Delta_+$ going down the page,
    interleaved with the map $(Rf)^\op : \ord{5} \to \ord{4}$ in $\Delta_=$ going up
    the page. The elements in $\Delta_+$ correspond to the gaps between elements
    in $\Delta_=$, so each map determines the other.
  }
\label{fig:delta-gaps}
\end{figure*}

\begin{figure*}[t]
$$\begin{tikzcd}[column sep = 0.3cm]
    X \ar{d}{f} &
    &X(\reg{0}) \ar{r} \ar{dl} &
    X(\sing{0}) \ar{dl} &
    X(\reg{1}) \ar{l} \ar{r} \ar{dl} \ar{dr} &
    X(\sing{1}) \ar{dr} &
    X(\reg{2}) \ar{l} \ar{r} &
    X(\sing{2}) \ar{dl} &
    X(\reg{3}) \ar{l} \ar{dr} \ar{dl} \\
    Y &
    Y(\reg{0}) \ar{r} &
    Y(\sing{0}) &
    Y(\reg{1}) \ar{l} \ar{r} &
    Y(\sing{1}) &
    Y(\reg{2}) \ar{l} \ar{r} &
    Y(\sing{2}) &
    Y(\reg{3}) \ar{l} \ar{r} &
    Y(\sing{3}) &
    Y(\reg{4}) \ar{l}
  \end{tikzcd}
$$

\vspace{-5pt}
\caption{A map $f: X \to Y$ of zigzags, with underlying singular and regular monotone maps as given in Figure~\ref{fig:delta-gaps}.}
\label{fig:zigzag-map}
\end{figure*}
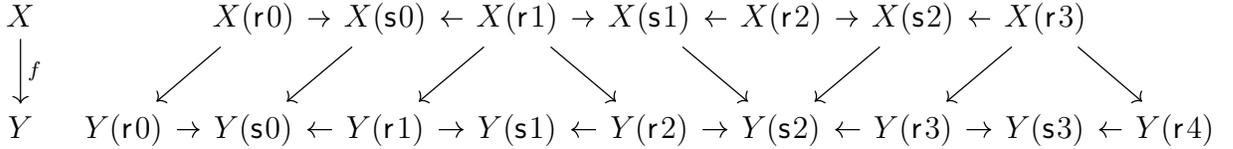

\begin{definition}\label{definition:zigzag}
  In a category \cat C, a \textit{zigzag} $X$ is a diagram of the following form, for some integer length $n \geq 0$:
  \[
    \begin{tikzcd}[column sep = 0.7cm]
      X(\reg{0}) \ar{r} & X(\sing{0}) & X(\reg{1}) \ar{l} \ar{r} & \cdots & \ar{l} X(\reg{n})
    \end{tikzcd}
  \]
The objects of the form $X(\reg{i})$ are called the \textit{regular
  objects}, and the objects of the form $X(\sing{i})$ the \textit{singular objects}. 
\end{definition} 

\noindent
A zigzag can also be thought of as a sequence of cospans, with adjacent cospans sharing a base object.

To define maps between zigzags, we first require an auxilliary observation. There is an equivalence $R : \Delta_+ \to \Delta_=^\op$, originally described  by Wraith~\cite{Wraith}, which sends an object $\ord{n}$ to $\ord{n + 1}$, and a map $f : \ord{n} \to \ord{m}$ to the opposite of the map $\ord{m + 1} \to \ord{n + 1}$ defined as follows:
\[
  i \mapsto \min (\{ j \in \ord{n} \mid f(j) \geq i \} \cup \{ n \}).
\]
While the formula may appear non-obvious, the idea is straightforward, and we illustrate it in Figure~\ref{fig:delta-gaps}, showing how the monotones $f$ and $(Rf)^\op$ are interleaved in a simple way.


\begin{definition}\label{definition:zigzag-map}
In a category \cat C, given zigzags $X,Y$ of length $n,m$ respectively, a \textit{zigzag map} $f : X \to Y$ consists of a \textit{singular map} $\singmap{f} : \ord{n} \to \ord{m}$ in $\Delta_+$, along with an implied \textit{regular map} $\regmap f := (R \singmap f)^\op : \ord {m+1} \to \ord {n+1}$, together with the following additional structure:
  \begin{enumerate}
    \item for every $0 \leq i \leq m$ a map in $\cat{C}$:
      \[f(\reg{i}): X(\reg{\regmap f(i))} \to Y(\reg{i})\]
    \item for every $0 \leq j < m$ a map in $\cat{C}$:
      \[f(\sing{j}): X(\sing{j}) \to Y(\sing{\singmap{f}(j)})\]
  \end{enumerate}
The maps $f(\reg{i})$ are called the \textit{regular slices}, and the maps $f(\sing{j})$ are the \textit{singular slices}. These maps must satisfy the following equations, for every $0 \leq i < m$:
  \begin{enumerate}
    \item If ${(\singmap{f})}^{-1}(i)$ is nonempty, with initial value $p$ and final value $q$, then the following diagrams must commute, for all $j$ with $p \leq j < q$:
      \[
        \begin{tikzcd}[column sep = 0.5cm]
          X(\reg{p}) \ar[r] \ar[d, "f(\reg{i})"] &
          X(\sing{p}) \ar[d, "f(\sing{p})"] \\
          Y(\reg{i}) \ar[r] & Y(\sing{i})
        \end{tikzcd}
        \quad
        \begin{tikzcd}[column sep = 0.5cm]
          X(\sing{q}) \ar[d, "f(\sing{q})"] &
          X(\reg{q + 1}) \ar[l] \ar[d, "f(\reg{i+ 1})"]
          \\
          Y(\sing{i}) &
          Y(\reg{i + 1}) \ar[l]
        \end{tikzcd}
      \]
      \[
        \begin{tikzcd}
          X(\sing{j}) \ar[r] \ar[dr, "f(\sing{j})", swap] &
          X(\reg{j + 1}) &
          X(\sing{j + 1}) \ar[l] \ar[dl, "f(\sing{j + 1})"] \\
          & Y(\sing{i})
        \end{tikzcd}
      \]
    \item If ${(\singmap{f})}^{-1}(i)$ is empty,  this diagram must commute:
      \[
        \begin{tikzcd}
          & X(\reg{\regmap{f}(i)})
          \ar[dl, "f(\reg{i})", swap]
          \ar[dr, "f(\reg{i + 1})"]
          \\
          Y(\reg{i}) \ar[r] &
          Y(\sing{i}) &
          Y(\reg{i + 1}) \ar[l]
        \end{tikzcd}
      \]
  \end{enumerate}
\end{definition}

The notion of zigzag map is geometrically natural, and best understood via example. We illustrate it in Figure~\ref{fig:zigzag-map}. The regular slices $f(\reg i)$ and singular slices $f(\sing j)$ are drawn vertically, with the zigzag structure of $X$ drawn above, and of $Y$\ drawn below. As a result we naturally obtain a categorical diagram comprising 7 squares, and the equational part of the zigzag map definition simply requires  these to  commute. In this example the singular monotone $\singmap{f} : \ord{3} \to \ord{4}$ is defined by $\singmap f (0)=1$, $\singmap f (1) = 2$ and $\singmap f (2) = 2$, while the regular monotone $\regmap f : \ord 5 \to \ord 4$ acts as $\regmap f(0)=0$, $\regmap f(1)=1$, $\regmap f(2) = 1$, $\regmap f(3) = 3$ and $\regmap f(4)=3$. 


Our approach slightly generalizes the original zigzag definition of Reutter and Vicary~\cite{Reutter_2019}, where the regular slices of a zigzag map were required to be identities. This extra generality will be critical for our results, since normalisation can change  the boundary of a diagram.

\suckpara
\paragraph{Categories of Zigzags}

  Zigzag maps can be composed in a natural way. Given $f:X \to Y$ and $g:Y \to W$, we define the following:
\begin{align*}
(g \circ f)(\sing j) &= g(\sing {\singmap f (j)})\circ f(\sing j)
\\
(g \circ f)(\reg i) &= g(\reg i) \circ f(\reg \regmap g(i))
\end{align*}
In terms of the representation used in Figure~\ref{fig:zigzag-map}, this corresponds to stacking one  diagram above the other. This is easily seen to be associative and unital, and hence for any category \cat C, we obtain a \textit{zigzag category} $Z(\cat{C})$ of zigzags and
  zigzag maps. This construction is functorial in $\cat{C}$.

The category $Z(1)$ of zigzags in the terminal category is isomorphic to
$\Delta_+$. For every category $\cat{C}$ the unique functor $\cat{C}
\to 1$ thus induces a functor $\pi : Z(\cat{C}) \to \Delta_+$, which is
natural in $\cat{C}$. In the next section we will make heavy use of the theory of cartesian and cocartesian lifts of maps in $\Delta_+$ to $Z(\cat{C})$, as they will allow us  to characterise universal zigzag maps of a particular shape.

Since for any category $\cat{C}$ the zigzag construction $Z(\cat{C})$ is a
category itself, the construction can be iterated. We write $Z^{n}(\cat{C})$
for the \textit{$n$-fold zigzag category} on $\cat{C}$.

\suckpara
\paragraph{Diagrams from Zigzags}

We define an \emph{n-diagram} to be an object of the category $Z^n(\N)$, where $\N$ denotes the poset of natural numbers $0< 1< \ldots$.  We consider such objects of $Z^n(\N)$ as combinatorial encodings of $n$\-dimensional string diagrams. To motivate this definition, we refer back to Figure~\ref{fig:zigzagmotivation}; using the machinery we have developed, we can now see that this represents an object of $Z^2(\N)$, giving a combinatorial foundation for the string diagram that appears onthe left-hand side of the figure.

The natural numbers at each point encode the dimension of the algebraic generator that exists at that location in the diagram. In a real string diagram, we would ordinarily give further information, labelling the points with the name of a generator. However, for the purposes of normalisation, only the dimensions of the generators are relevant, and so this simpler notation suffices for our purposes.

To represent a meaningful string diagram,  an $n$\-diagram  must also satisfy \textit{type-checking} conditions, for which normalisation plays a critical role. We describe this in Section~\ref{sec:implementation}.


\sucksec
\section{Degeneracy Maps}
The zigzag construction admits a notion of \textit{degeneracy map} that insert ``identity
regions'' into a diagram. A degeneracy map $X \to Y$ then serves as a witness that the diagrams 
$X$ and $Y$ are similar, with the only difference being that $X$ contains fewer redundant identities. In this section we define degeneracy maps and study their properties. Significant use is made of cartesian and co-cartesian liftings.

\suckpara
\paragraph{Cartesian Liftings}
For some $\ord{n}$ in $\Delta_+$ the $i$th \textit{face map} is the unique injective map $d_i: \ord{n} \to \ord{n + 1}$ that omits $i \in
\ord{n + 1}$ from its image. We can illustrate this as follows:
$$
\begin{tikzpicture}[xscale=1.5]
\node at (-1,0) {$\ldots$};
\node (1) at (0.,0) {$i-1$};
\node (2) at (1.,0) {$i$};
\node (3) at (2.,0) {$i+1$};
\node at (3.5,0) {$\ldots$};
\node at (-1,-1) {$\ldots$};
\node (4) at (0,-1) {$i-1$};
\node (5) at (1,-1) {$i$};
\node (6) at (2,-1) {$i+1$};
\node (7) at (3,-1) {$i+2$};
\node at (4,-1) {$\ldots$};
\draw [->] (1) to (4);
\draw [->] (2) to (5);
\draw [->] (3) to (7);
\end{tikzpicture}
$$
We will use these face maps to construct zigzag maps with an important categorical lifting property, as follows.

\suckpara
\begin{definition}
  Let $p : \cat{C} \to \cat{D}$ be a functor. A map \mbox{$f : x \to y$} in $\cat{C}$ is
  \emph{$p$-cartesian} if for every map $h : x' \to y$ in $\cat{C}$ and $u : p(x') \to p(y)$
  such that $p(h) = p(f) \circ u$, there exists a unique $v : x' \to x$ in \cat C such that
  $h = f \circ v$ and $u = p(v)$:
  \[
    \begin{tikzcd}
      x' \ar[dr, "h"] \ar[d, "v", swap, dashed] \\[-5pt]
      x \ar[r, swap, "f"] & y
    \end{tikzcd}
    \quad
    \mapsto _p
    \quad
    \begin{tikzcd}
      p(x') \ar[dr, "p(h)"] \ar[d, "u", swap] \\[-5pt]
      p(x) \ar[r, swap, "p(f)"] & p(y)
    \end{tikzcd}
  \]
The \emph{$p$-vertical} maps are those maps $f : x \to y$ such that $p(f) =
  \id$.  A map is \textit{$p$-cocartesian} if it is $p^\op$-cartesian.
\end{definition}

\noindent
This is widely used property in categorical algebra, with a simple intuition: one imagines that $f,h$ are paths in a space, with a projection $p$ to some subspace, such that whenever $h$ factors through $f$ in the projection, the factorisation can  be lifted to the original space.

Cartesian maps are a standard concept from the theory of fibred categories~\cite{SGA1, vistoli2007notes}.
The functors that we consider in this paper will not be categorical fibrations or opfibrations,
but nevertheless, certain maps will still satisfy the universal property of cartesian or cocartesian maps, as we now show.


\begin{lemma}\label{lemma:degeneracy-cocartesian}
In a category \cat C, for any zigzag $A$ of length $n$ and $i \in \ord{n+1}$, the following zigzag map out of $A$ is $\pi$\-cocartesian over the face map $d_i$:
\[
\label{eq:simple-degeneracy}
    \begin{tikzcd}[column sep=15pt]
      \cdots & & \ar{ll} \ar[swap]{dl}{\id} A(\reg{i}) \ar{dr}{\id} \ar{rr} & & \cdots \\
      \cdots & \ar{l} A(\reg{i}) \ar[swap]{r}{\id} & A(\reg{i}) & \ar{l}{\id} A(\reg{i}) \ar{r} & \cdots
    \end{tikzcd}
\]
Here the singular map is the $i$th face map, and we insert an itentity cospan  in the target, with the slice maps also being identities
\end{lemma}
\begin{proof}
  Let $f : A \to A'$ be a map of this form. Let $h : A \to B$ be a map in
  $Z(\cat{C})$ and $u : \pi(A') \to \pi(B)$ a map in $\Delta_+$
  such that $u \circ \pi(f) = \pi(h)$. The shape of the diagram
  defining a lift $v : A' \to B$ of $u$ is completely determined. Since the slices of $f$ 
  and the cospan inserted in $A'$ are identities, the slices of $v$ are
  completely determined by the requirement that \mbox{$v \circ f = h$}.
\end{proof}

\suckpara
\paragraph{Generating Degeneracies}
In the category $\Delta_+$, the face maps generate all the monomorphisms.
Since $\pi$-cocartesian maps are unique up to unique vertical isomorphism~\cite{vistoli2007notes}, this lemma therefore implies
that any $\pi$-cocartesian map over a monomorphism in $\Delta_+$ inserts levels consisting
of isomorphisms. We call those maps the \textit{simple degeneracy maps}.

Diagrams of higher dimension admit more ways to insert identities: not only can
we insert a $1$-dimensional identity slice into a $2$-dimensional diagram, but
we can degenerate each of the $1$-dimensional subslices. We call these zigzag
maps $f$ with $\pi(f) = \id$ that have degeneracy maps as slices the
\textit{parallel degeneracy maps}.
We represent higher-dimen-sional diagrams by iterating the zigzag construction, so we
can define general degeneracy maps recursively, as follows.

\begin{definition}\label{def:degeneracy-map}
  \textit{Degeneracy maps} in $Z^{n}(\cat{C})$ are generated under composition by the following classes:
  \begin{enumerate}
    \item \textit{simple degeneracy maps}, the $\pi$-cocartesian maps over the monomorphisms in $\Delta_+$;
    \item \textit{parallel degeneracy maps}, the $\pi$-vertical maps in which every slice map is a degeneracy map in $Z^{n - 1}(\cat{C})$.
  \end{enumerate}
\end{definition}

To simplify inductive arguments we define degeneracy maps in $Z^0(\cat{C})$ to
be the isomorphisms.

\begin{lemma}\label{lemma:degeneracy-isomorphism}
Isomorphisms in $Z^n(\cat{C})$ are degeneracy maps.
\end{lemma}
\begin{proof}
Let $f : A \to B$ be an isomorphism in $Z^n(\cat{C})$. Since $\pi$ preserves
isomorphisms and $\Delta_+$ is skeletal, $f$ is $\pi$-vertical. Since the slice
maps of an isomorphism in $Z^n(\cat{C})$ need to be isomorphisms in $Z^{n - 1}(\cat{C})$,
by induction they are degeneracy maps as well. Hence $f$ is a parallel degeneracy map.
\end{proof}


\begin{lemma}\label{lemma:degeneracy-decomposition}
  Let $f : A \to B$ be a degeneracy map in $Z^{n}(\cat{C})$. Then $f$ factors
  uniquely (up to isomorphism) into a simple degeneracy map
  followed by a parallel degeneracy map.
\end{lemma}
\begin{proof}
  Since the maps in Definition~\ref{def:degeneracy-map} are sent by $\pi$ to
  either a monomorphism or an identity map, we have that $\pi(f)$ is a
  monomorphism as well. By Lemma~\ref{lemma:degeneracy-cocartesian} the map \mbox{$f_1 : A
  \to A'$} of the shape $\pi(f)$ which inserts identity levels is
  $\pi$-cocart-esian, so there exists a unique $\pi$-vertical map $f_2 : A' \to B$
  such that $f = f_2 \circ f_1$.
  Since the slice maps of the maps in Definition~\ref{def:degeneracy-map} are either
  isomorphisms or degeneracy maps in $Z^n(\cat{C})$, the slice maps of $f$ must be degeneracy maps.
  But
  the slice maps of $f_1$ are identities, so the slices of $f_2$ must be degeneracy maps as well.
\end{proof}

\suckpara
\paragraph{Degeneracy Maps as Subobjects.}

Any given  degeneracy map \mbox{$f : X \to Y$} can be interpreted as a witness that $X$ as a subobject of $Y$, which omits some identity regions. This intuition is reflected in the theory as follows.

\begin{lemma}\label{lemma:degeneracy-monomorphism}
  Degeneracy maps in $Z^n(\cat{C})$ are monomorphisms.
\end{lemma}
\begin{proof}
  Since monomorphisms are closed under composition it suffices to prove the
  claim for the generating maps of Definition~\ref{def:degeneracy-map}:

  \begin{itemize}
    \item Let $f : X \to Y$ be a simple degeneracy map, and let \mbox{$g_1, g_2 : Z \to X$} be maps such that \mbox{$f \circ g_1 = f \circ g_2$}. Since $\pi(f)$ is a monomorphism
      we have $\pi(g_1) = \pi(g_2)$. The slices of $f$ are isomorphisms, hence $g_1$
      and $g_2$ must have equal slices, and so $g_1 = g_2$. Thus $f$ is a monomorphism.

    \item Now let $f : X \to Y$ be a parallel degeneracy map and $g_1, g_2 : Z \to X$ a pair
      of maps such that $f \circ g_1 = f \circ g_2$. Since $\pi(f) = \id$ 
      we have $\pi(g_1) = \pi(g_2)$. The slices of $f$ are degeneracies, so by
      induction they are monomorphisms. Hence $g_1$ and $g_2$ must have equal slices
      and so $g_1 = g_2$. Thus $f$ is a monomorphism as well.\qedhere
  \end{itemize}
\end{proof}

\begin{example}
  We note  that the converse of the previous lemma does not hold: not every monomorphism is a degeneracy map.   The following map of zigzags is a monomorphism in $Z(\Delta_+)=Z^2(1)$, but not a degeneracy map:

  \[
    \begin{tikzcd}
      & \ord{2} \ar[swap, dl, "\id"] \ar[dr, "\id"] \\[-10pt]
      \ord{2} \ar{r} & \ord{1} & \ord{2} \ar{l}
    \end{tikzcd}
  \]

\end{example}

Just like monomorphisms, degeneracy maps satisfy the following closure property, which we establish with a simple inductive argument.

\begin{lemma}\label{lemma:degeneracy-closure}
  For any commutative triangle in $Z^n(\cat{C})$ as follows, if $f,g$ are degeneracy maps, so is $\varphi$:
  \[
    \begin{tikzcd}
      A \ar[swap]{d}{\varphi} \ar{r}{f} & T \\
      B \ar[swap]{ur}{g}
    \end{tikzcd}
  \]
\end{lemma}
\begin{proof}
  For $n = 0$ this follows since isomorphisms satisfy $2$-out-of-$3$. For $n >
  0$ we have that $\pi(f)$ and $\pi(g)$ are monomorphisms and hence
  $\pi(\varphi)$ must be a monomorphism as well. Consider the diagram obtained
  by gluing together the defining diagrams $f$, $g$ and $\varphi$. Then every
  slice map of $\varphi$ is contained in a commutative triangle of the form
  above, so by induction is a degeneracy map.
\end{proof}

By Lemma~\ref{lemma:degeneracy-monomorphism} a degeneracy map $X \to T$ is a
monomorphism, and thus represents a subobject of $T$. Two monomorphisms $f : X \to T$
and $g : X \to T$ represent the same subjobject of $T$ when there exists an isomorphism
$\varphi : X \to Y$ such that $g \circ \varphi = f$. Since by Lemma~\ref{lemma:degeneracy-isomorphism} all isomorphisms are degeneracy
maps and degeneracy maps are closed under composition, if one monomorphism representing
some subobject of $T$ is a degeneracy map, then all of them are.
This allows us to define the subposet $\Deg(T) \subseteq \Sub(T)$
of degeneracies into~$T$.

\begin{definition}
  Let $\cat{C}$ be a category and $n \geq 1$. For $T \in Z^n(\cat{C})$ let
  $\Deg(T)$ be the subposet of $\Sub(T)$ consisting of those subobjects of $T$
  represented by a degeneracy map into~$T$.
\end{definition}

\sucksec
\section{Diagram normalisation}

The normalisation of an object $T$ of $Z^n(\cat{C})$ for some \mbox{$n \geq 0$} is the smallest
element of $\Deg(T)$, if it exists, in which  all redundant identities have been removed.
In this section we show that normalisations exist, and describe them as the meet of all of the elements of $\Deg(T)$. Meets in $\Sub(T)$ are intersections of subobjects,
which are calculated by taking the pullback of representatives.

\begin{lemma}\label{lemma:degeneracy-cartesian}
A morphism of zigzags $A\to B$ in $Z(\cat{C})$ over the $i$th face map $d_i: \ord{n} \to \ord{n+1}$ in $\Delta_+$ is $\pi$\-cartesian if and only if the morphisms in Figure~\ref{fig:cartesian} indicated by $\cong$ are isomorphisms in $\cat{C}$, and the square indicated by $\lrcorner$ is a pullback square in $\cat{C}$.
\end{lemma}
\begin{proof}
  Let $f : A \to B$ be a map of this form. Let $h : A' \to B$ be a map of
  $Z(\cat{C})$ and $u : \pi(A') \to \pi(A)$ a map of $\Delta_+$ such that
  $\pi(f) \circ u = \pi(h)$. We need to construct a lift $v : A' \to A$ of $u$
  that satisfies $f \circ v = h$. The slice $v(\reg{i})$ is uniquely determined by the
  universal property of the pullback square in the defining diagram of $f$. The other
  slices of $v$ are determined since the slices of $f$ to the left and right of the
  pullback square are isomorphisms. The converse follows by essential uniqueness of $\pi$\-cartesian maps.
\end{proof}

\begin{figure*}[t!]
$$\begin{tikzcd}[column sep = 0.3cm]
        {A(\reg{0})} & {A(\sing{i-1})} && {A(\reg{i})} && {A(\sing{i})} & {A(\reg{n})} \\
        {B(\reg{0})} & {B(\sing{i-1})} & {B(\reg{i})} & {B(\sing{i})} & {B(\reg{i+1})} & {B(\sing{i+1})} & {B(\reg{n+1})}
        \arrow[from=1-4, to=1-2]
        \arrow[from=1-4, to=1-6]
        \arrow["\cong", from=1-1, to=2-1]
        \arrow["\cong", from=1-2, to=2-2]
        \arrow[from=2-3, to=2-2]
        \arrow[from=2-3, to=2-4]
        \arrow[from=2-5, to=2-4]
        \arrow[from=1-4, to=2-3]
        \arrow[from=1-4, to=2-5]
        \arrow["\lrcorner"{anchor=center, pos=0.125, rotate=-45}, draw=none, from=1-4, to=2-4]
        \arrow[dotted, no head, from=1-1, to=1-2]
        \arrow[dotted, no head, from=2-1, to=2-2]
        \arrow[dotted, no head, from=1-6, to=1-7]
        \arrow[dotted, no head, from=2-6, to=2-7]
        \arrow["\cong", from=1-6, to=2-6]
        \arrow["\cong", from=1-7, to=2-7]
        \arrow[from=2-5, to=2-6]
\end{tikzcd}$$

\vspace{-8pt}
\caption{\label{fig:cartesian}A $\pi$-cartesian map of zigzags over the $i$th face map $d_i:\ord{n} \to\ord{n+1}$.}
\end{figure*}


In particular, a square consisting of isomorphisms is a pullback square, so
all simple degeneracy maps are $\pi$-cart-esian maps. Using the machinery of cartesian lifts and the following proposition, we can prove that $\Deg(T)$ is closed under intersections. These intersections are a rare instance of limits which exist in $Z^n(\cat{C})$, independently of the existence of limits in $\cat{C}$. 

\begin{proposition}\label{lemma:degeneracy-pullback}
For any category \cat C, for any  two degeneracy maps \mbox{$f : X \to T$} and $g : Y \to T$ in
  $Z^n(\cat{C})$, their pullback exists, and the projections are also degeneracy maps.
  In particular $\Deg(T)$ is closed under intersection of subobjects.
\end{proposition}
\begin{proof}
  For $n = 0$, the pullback of a pair of isomorphisms exists and
  the projections are isomorphisms again. We now proceed to the case $n > 0$ by
  induction.
  There is a full embedding of $Z^n(\cat{C})$ into $Z(\PSh{Z^{n - 1}(\cat{C})})$ induced by
  the Yoneda embedding. We first calculate the pullback there and afterwards
  show that it consists of representable objects with degeneracy maps as
  projectors from the pullback. The claim then follows because full embeddings reflect limits.

  Since $f$ and $g$ are degeneracy maps, the induced maps $\pi(f)$ and $\pi(g)$
  are monomorphisms. Pullbacks of monomorphisms exist in $\Delta_+$ and are monomorphisms
  themselves, so we have a pullback square in which every map is a monomorphism:
  \[
    \begin{tikzcd}
      I \ar{r} \ar{d} & \pi(B) \ar{d} \\[-7pt]
      \pi(A) \ar{r} & \pi(T)
    \end{tikzcd}
  \]
Since the presheaf category $\PSh{Z^{n - 1}(\cat{C})}$ is complete, the pullbacks necessary to apply Lemma~\ref{lemma:degeneracy-cartesian} exist, and so there are $\pi$-cartesian lifts of the maps from $I$ as well as unique $\pi$-vertical maps between them that make the following diagram commute:
  \[
    \begin{tikzcd}[column sep = 0.5cm]
      & B' \ar{d} \ar{dr} \\[-9pt]
      A' \ar{r} \ar{dr} & T' \ar{dr} & B \ar{d}  \\[-9pt]
      & A \ar{r} & T
    \end{tikzcd}
    \quad\mapsto_\pi\quad
    \begin{tikzcd}[column sep = 0.5cm]
      & I \ar[equal]{d} \ar{dr} \\[-9pt]
      I \ar[equal]{r} \ar{dr} & I \ar{dr} & \pi(B) \ar{d}  \\[-9pt]
      & \pi(A) \ar{r} & \pi(T)
    \end{tikzcd}
  \]
  Then the pullback of the cospan $A' \to T' \leftarrow B'$ in the fibre over $I$ is the pullback
  of the original cospan $A \to T \leftarrow B$:
  \[
    \begin{tikzcd}
      P \ar{r} \ar{d} & B' \ar{d} \ar{dr} \\[-9pt]
      A' \ar{r} \ar{dr} & T' \ar{dr} & B \ar{d}  \\[-9pt]
      & A \ar{r} & T
    \end{tikzcd}
  \]
  The fibres of $\pi$ are diagram categories in which limits are determined
  pointwise. We thus have that together with Lemma~\ref{lemma:degeneracy-cartesian}
  the objects of $P$ are limits of diagrams of the following form, where every horizontal map is a degeneracy map, and at least one of the outer columns consists of isomorphisms:
  \[
    \begin{tikzcd}
      A_0 \ar{r} \ar{d} & T_0 \ar{d} & B_0 \ar{d} \ar{l} \\[-8pt]
      A_1 \ar{r} & T_1 & B_1 \ar{l} \\[-8pt]
      A_2 \ar{r} \ar{u} & T_2 \ar{u} & B_2 \ar{u} \ar{l}
    \end{tikzcd}
  \]
   Without loss of generality we can assume that it is the left-most
  column and the isomorphisms are identities. By induction, the pullbacks
  $P_0$, $P_1$ and $P_2$ of the rows are representable, and the projection maps
  are degeneracy maps, so we get a diagram as follows, in which every row is a degeneracy map:
  \[
    \begin{tikzcd}
      P_0 \ar{r} \ar{d} & A_0 \ar[equal]{d} \\[-8pt]
      P_1 \ar{r} & A_1 \\[-8pt]
      P_2 \ar{r} \ar{u} & A_2 \ar[equal]{u}
    \end{tikzcd}
  \]
  By Lemma~\ref{lemma:degeneracy-closure}
  the vertical maps are degeneracy maps as well and their pullback is representable
  by induction. Hence every object in the pullback in $Z(\PSh{Z^{n - 1}(\cat{C})})$ is
  representable and the slice maps of the projections are are degeneracy maps
  as required.
\end{proof}

While this lemma proves that $\Deg(T)$ is closed under binary and hence finite
intersections, we need the intersection of all elements in $\Deg(T)$. Since
degeneracy maps are uniquely determined up to isomorphism by their action on
the shape of a diagram, and any shape only has a finite number of identity
regions to be removed, finite intersections will be enough.

\begin{lemma}\label{lemma:degeneracy-finite}
  For any two degeneracy maps $f : X \to T$ and $g : Y \to T$ in $Z^n(\cat C)$ that are sent to the
  same map of untyped diagrams in $Z^n(1)$, there exists an isomorphism $\varphi : X \to Y$
  such that $f = g \circ \varphi$. In particular $\Deg(T)$ is finite.
\end{lemma}
\begin{proof}
  For $n = 0$ the degeneracy maps are isomorphisms. For $n \geq 1$ we factor
  degeneracy maps into top-level degeneracy maps followed by a parallel
  degeneracy map. 
  \[
    \begin{tikzcd}
      X \ar{r} \ar[dashed]{dd} & X' \ar{dr} \ar[dashed]{dd} \\[-18pt]
      & & T \\[-18pt]
      Y \ar{r} & Y' \ar{ur}
    \end{tikzcd}
  \]
  We now construct isomorphisms $X' \to Y'$ and $X \to Y$ that fit into this
  diagram as follows.  By induction there exist isomorphisms between the slices of $X'$ and $Y'$
  that commute with the cospans to form an isomorphism $X' \to Y'$ since
  the slice maps of $Y' \to T$ are monomorphisms by Lemma~\ref{lemma:degeneracy-monomorphism}.
  Now by Lemma~\ref{lemma:degeneracy-cartesian} the maps $X \to X'$ and $Y \to Y'$
  are $\pi$\-cartesian. Since they are sent to the same map in $Z(1)$ by~$\pi$,
  there exists an isomorphism $X \to Y$ that makes the diagram commute.
\end{proof}

We can thus conclude that any object in an iterated zigzag category admits a normalisation.

\begin{proposition}\label{lemma:degeneracy-initial}
  For any $T \in Z^n(\cat{C})$ the poset $\Deg(T)$ has a smallest element.
\end{proposition}

\begin{proof}
  $\Deg(T)$ is finite by Lemma~\ref{lemma:degeneracy-finite}, so the binary
  intersections of Proposition~\ref{lemma:degeneracy-pullback} suffice to construct
  the intersection of all elements of $\Deg(T)$, yielding the smallest element.
\end{proof}

\noindent
Let $T \in Z^n(\cat{C})$, and suppose $n : N \to T$ is a degeneracy map representing
the smallest element of $\Deg(T)$. Then $n$ is called a \textit{normalising map},
and $N$ is the \textit{normalisation} of $T$.

\sucksec
\section{Computing Normalisation}
\label{sec:computing}

\paragraph{Essential Identities.}
In order to be implemented as part of a type checking algorithm, we need a way to compute the normalisation of a diagram. 
As a first attempt, we might consider a naive recursive algorithm, where we normalise all the singular and regular objects of a zigzag, levelwise. 

However, this cannot work. To see why, consider the diagram in
Figure~\ref{fig:normalisation-sink}, in which the top zigzag $T$ and bottom zigzag $B$ are normalised, but  the middle zigzag  $M$ is not. If we normalised the middle row, we would obtain a new zigzag $M'$ with length $0$,
and then the updated zigzag map \mbox{$q':T \to M'$} from the top row to the middle row
would require a singular map of type of $\ord{2} \to \ord{0}$ in $\Delta_+$. But there are no such functions, as $\ord 0$ is empty.

As a result, the identity cospan contained in $M$ is not redundant, but \textit{essential} for the geometry of the entire structure, and cannot be removed. The entire structure shown in Figure~\ref{fig:normalisation-sink} is therefore \textit{already normalised}, despite the existence of the identity cospan in the middle row.

What makes the normalisation algorithm nontrivial is that it must correctly detect these essential identities, leaving them in place, while removing the redundant identities. Note that the original definition of normalisation via Proposit-ion~\ref{lemma:degeneracy-initial} handles this subtlety automatically in some sense, since an essential identity cannot be factorized out. But for our normalisation algorithm, we must handle it explicitly.

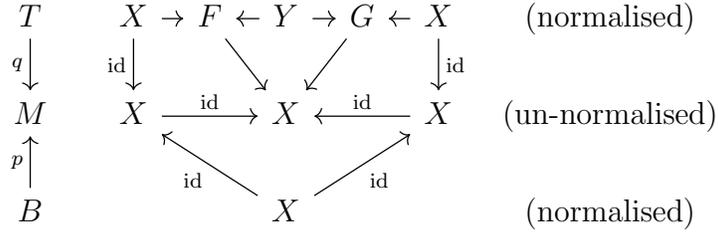
\begin{figure}[t]
  \[
    \begin{tikzcd}[column sep = 0.3cm]
      T \ar[swap]{d}{q} &&X \ar{r} \ar[swap]{d}{\id} & F \ar{dr} & Y \ar{l} \ar{r} & G \ar{dl} & X \ar{l} \ar{d}{\id} & \text{(normalised)}\\[-5pt]
      M&&X \ar{rr}{\id} && X && X \ar[swap]{ll}{\id} & \text{(un-normalised)} \\[-5pt]
      B \ar{u}{p}&&&& X \ar{ull}{\id} \ar[swap]{urr}{\id} &&& \text{(normalised)}
    \end{tikzcd}
  \]
  
  \vspace{-8pt}
  \caption{A normalised diagram with an un-normalised level.}\label{fig:normalisation-sink}
\end{figure}


\suckpara
\paragraph{Sink Normalisation}

To solve this problem, instead of normalising each part of a diagram in isolation, we
keep track of that part's ``environment'', in the form of a \textit{sink} of incoming maps $(T,\{f_i : A_i \to T \}_{i \in I})$, where $I$ is an indexing set, which we will often omit when it is clear from context. In the case of Figure~\ref{fig:normalisation-sink}, we would
ask for a normalisation of the middle zigzag $M$ in the context of the sink $$(M,\{\mbox{$p:B \to M$}, \mbox{$\,q:T \to M$}\})$$ with two incoming zigzag maps.

Proposition~\ref{lemma:degeneracy-initial} showed that for any \mbox{$T\in Z^n(\cat{C})$}, the poset $\Deg(T)$ has a smallest element $N \to T$, which we defined as the normalisation of $T$. In the following, we prove a relative version of this proposition, yielding a notion of normalisation \emph{relative} to a sink. Given a sink $\mathcal{S} = (T,\{f_i: A_i \to T\})$, let $\Deg_{\mathcal{S}}(T)$ denote a subposet of $\Deg(T)$, containing those degeneracy maps $n: N\hookrightarrow T$ through which the sink $\mathcal{S}$ factors, i.e. such that there exists $(N,\{g_i : A_i \to N\})$ such that $f_i = n \circ g_i$. Intuitively, the idea is that $N$ is a subobject of $T$ that arises by discarding only those redundant identities which are \textit{not} in the image of an element of the sink $f_i$.

\begin{proposition}\label{lemma:initialdegeneratesink}
  Let $\mathcal{S} = (T,\{ f_i: A_i \to T \})$ be a sink of maps in $Z^n(\cat{C})$. Then the
  subposet $\Deg_{\mathcal{S}}(T)$ of the finite poset $\Deg(T)$ is non-empty and   closed under intersection, and therefore also has a smallest element.  
\end{proposition}
\begin{proof}
The identity $\id: T\to T$ is a degeneracy map through which every sink factors, and hence is an element of $\Deg_{\mathcal{S}}(T)$. 
  The intersection of subobjects is the pullback of representatives, which
  exists for degeneracy maps due to Proposition~\ref{lemma:degeneracy-pullback}.
  The factorisations through the intersections are then given by the universal
  property of the pullback. By the same argument as in the proof of Proposition~\ref{lemma:degeneracy-initial}, it follows that $\Deg_{\mathcal{S}}(T)$ has a smallest element. 
\end{proof}

We call the initial element of $\Deg_{\mathcal{S}}(T)$ the \textit{relative normalisation} of $T$ with respect to the sink $\mathcal{S}$. If $\mathcal{S}$ is the empty sink, we have $\Deg_{\mathcal{S}}(T) = \Deg(T)$, and hence the relative normalisation of $T$ with respect to the empty sink $(T,\emptyset)$ agrees with the normalisation of $T$. In this way, we see that full normalisation is a special case of relative normalisation. The reason to study relative normalisation is that it admits a recursive algorithm, as follows.

  \begin{figure*}[t!]
  $
  \begin{tikzpicture}[xscale=.7, yscale=.75, font=\scriptsize]
  \node (ar0) at (0,0) {$A(\reg 0)$};
  \node (as0) at (0,1) {$A(\sing 0)$};
  \node (ar1) at (0,2) {$A(\reg 1)$};
  \node (as1) at (0,3) {$A(\sing 1)$};
  \node (ar2) at (0,4) {$A(\reg 2)$};
  \node (as2) at (0,5) {$A(\sing 2)$};
  \node (ar3) at (0,6) {$A(\reg 3)$};
  \draw [->] (ar0) to (as0);
  \draw [->] (ar1) to (as0);
  \draw [->] (ar1) to (as1);
  \draw [->] (ar2) to (as1);
  \draw [->] (ar2) to (as2);
  \draw [->] (ar3) to (as2);

  \node (tr0) at (4,0) {$T(\reg 0)$};
  \node (ts0) at (4,1) {$T(\sing 0)$};
  \node (tr1) at (4,2) {$T(\reg 1)$};
  \node (ts1) at (4,3) {$T(\sing 1)$};
  \node (tr2) at (4,4) {$T(\reg 2)$};
  \node (ts2) at (4,5) {$T(\sing 2)$};
  \node (tr3) at (4,6) {$T(\reg 3)$};
  \draw [->] (tr0) to (ts0);
  \draw [->] (tr1) to (ts0);
  \draw [->] (tr1) to (ts1);
  \draw [->] (tr2) to (ts1);
  \draw [->] (tr2) to (ts2);
  \draw [->] (tr3) to (ts2);
  
  \draw [->] (ar0) to node [above] {$f(\reg 0)$} (tr0);
  \draw [->] (as0) to node [above] {$f(\sing 0)$} (ts0);
  \draw [->] (as1) to node [above] {$f(\sing 1)$} (ts0);
  \draw [->] (ar2) to node [above] {$f(\reg 1)$} (tr1);
  \draw [->] (ar2) to node [above] {$f(\reg 2)$} (tr2);
  \draw [->] (as2) to node [above] {$f(\sing 2)$} (ts2);
  \draw [->] (ar3) to node [above] {$f(\reg 3)$} (tr3);
  
  \node (A) at (0,-1) {$A$};
  \node (T) at (4,-1) {$T$};
  \draw [->] (A) to node [below] {$f$} (T);

  \end{tikzpicture}
\hspace{2cm}
  \begin{tikzpicture}[xscale=.7, yscale=.75, font=\scriptsize]
  \node (ar0) at (0,0) {$A(\reg 0)$};
  \node (as0) at (0,1) {$A(\sing 0)$};
  \node (ar1) at (0,2) {$A(\reg 1)$};
  \node (as1) at (0,3) {$A(\sing 1)$};
  \node (ar2) at (0,4) {$A(\reg 2)$};
  \node (as2) at (0,5) {$A(\sing 2)$};
  \node (ar3) at (0,6) {$A(\reg 3)$};
  \draw [->] (ar0) to (as0);
  \draw [->] (ar1) to (as0);
  \draw [->] (ar1) to (as1);
  \draw [->] (ar2) to (as1);
  \draw [->] (ar2) to (as2);
  \draw [->] (ar3) to (as2);

  \node (tr0) at (12,0) {$T(\reg 0)$};
  \node (ts0) at (12,1) {$T(\sing 0)$};
  \node (tr1) at (12,2) {$T(\reg 1)$};
  \node (ts1) at (12,3) {$T(\sing 1)$};
  \node (tr2) at (12,4) {$T(\reg 2)$};
  \node (ts2) at (12,5) {$T(\sing 2)$};
  \node (tr3) at (12,6) {$T(\reg 3)$};
  \draw [->] (tr0) to node [right] {$t_0$} (ts0);
  \draw [->] (tr1) to node [right] {$t_0'$} (ts0);
  \draw [->] (tr1) to node [right] {$t_1$} (ts1);
  \draw [->] (tr2) to node [right] {$t_1'$} (ts1);
  \draw [->] (tr2) to node [right] {$t_2$} (ts2);
  \draw [->] (tr3) to node [right] {$t_2'$} (ts2);

  \node (pr0) at (8,0) {$P(\reg 0)$};
  \node (ps0) at (8,1) {$P(\sing 0)$};
  \node (pr1) at (8,2) {$P(\reg 1)$};
  \node (ps1) at (8,3) {$P(\sing 1)$};
  \node (pr2) at (8,4) {$P(\reg 2)$};
  \node (ps2) at (8,5) {$P(\sing 2)$};
  \node (pr3) at (8,6) {$P(\reg 3)$};
  \draw [->] (pr0) to node [right] {$p_0$} (ps0);
  \draw [->] (pr1) to node [right] {$p_0'$} (ps0);
  \draw [->] (pr1) to node [right] {$p_1{=}\id$} (ps1);
  \draw [->] (pr2) to node [right] {$p_1'{=}\id$} (ps1);
  \draw [->] (pr2) to node [right] {$p_2{=}\id$} (ps2);
  \draw [->] (pr3) to node [right] {$p_2'{=}\id$} (ps2);

  \node (nr0) at (4,1) {$P(\reg 0)$};
  \node (ns0) at (4,2) {$P(\sing 0)$};
  \node (nr1) at (4,3) {$P(\reg 1)$};
  \node (ns1) at (4,4) {$P(\sing 2)$};
  \node (nr2) at (4,5) {$P(\reg 3)$};
  \draw [->] (nr0) to node [right] {$p_0$} (ns0);
  \draw [->] (nr1) to node [right] {$p_0'$} (ns0);
  \draw [->] (nr1) to node [right] {$p_2$} (ns1);
  \draw [->] (nr2) to node [right] {$p_2'$} (ns1);

  \draw [->] (ar0) to (nr0);
  \draw [->] (as0) to (ns0);
  \draw [->] (as1) to (ns0);
  \draw [->] (ar2) to (nr1);
  \draw [->] (ar2) to (nr1);
  \draw [->] (as2) to (ns1);
  \draw [->] (ar3) to (nr2);

  \draw [->] (nr0) to node [above] {\id} (pr0);
  \draw [->] (ns0) to node [above] {\id} (ps0);
  \draw [->] (nr1) to node [above] {\id} (pr1);
  \draw [->] (nr1) to node [above] {\id} (pr2);
  \draw [->] (ns1) to node [above] {\id} (ps2);
  \draw [->] (nr2) to node [above] {\id} (pr3);

  \draw [->] (pr0) to node [above] {$q(\reg 0)$} (tr0);
  \draw [->] (ps0) to node [above] {$q(\sing 0)$} (ts0);
  \draw [->] (pr1) to node [above] {$q(\reg 1)$} (tr1);
  \draw [->] (ps1) to node [above] {$q(\sing 1)$} (ts1);
  \draw [->] (pr2) to node [above] {$q(\reg 2)$} (tr2);
  \draw [->] (ps2) to node [above] {$q(\sing 2)$} (ts2);
  \draw [->] (pr3) to node [above] {$q(\reg 3)$} (tr3);

  \node (A) at (0,-1) {$A$};
  \node (N) at (4,-1) {$N$};
  \node (P) at (8,-1) {$P$};
  \node (T) at (12,-1) {$T$};
  \draw [->] (A) to node [below] {$g$} (N);
  \draw [->] (N) to node [below] {$d_S$} (P);
  \draw [->] (P) to node [below] {$d_P$} (T);

  \end{tikzpicture}
  $

\vspace{-5pt}
\caption{\label{fig:algorithm_illustration}%
An illustration of the normalisation algorithm.}
  \end{figure*}
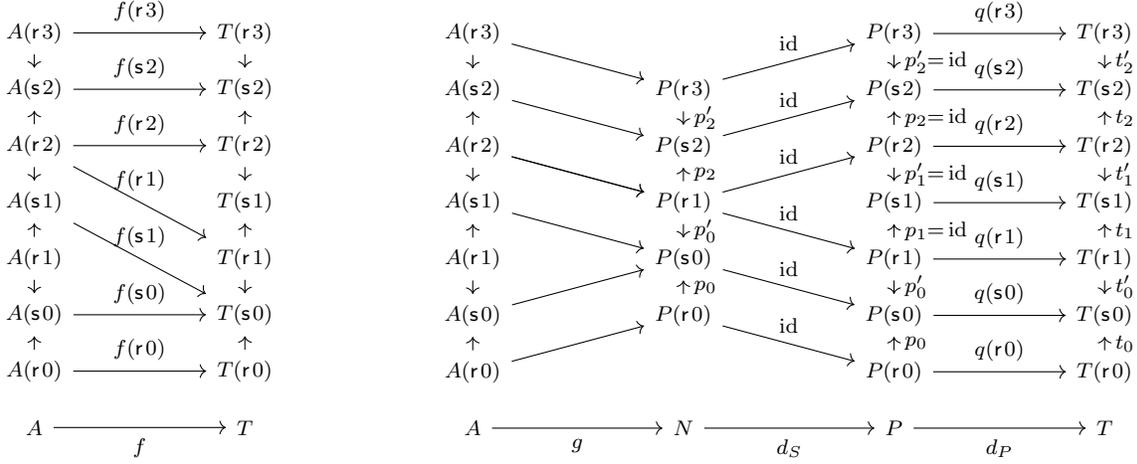

\paragraph{The Normalisation Algorithm.}
Given a sink in $Z^n(\cat{C})$, the following recursive algorithm computes its relative normalisation. We give a step-by-step illustration in Example~\ref{example:algorithm}.  
{\color{white}\begin{proposition}\label{constr:normalisation}Blah blah\end{proposition}}
\vspace{-15pt}

\noindent
\textbf{Construction~\ref*{constr:normalisation}} (Normalisation). Given a sink of maps $\mathcal{S} = (T,{\{f_i : A_i \to T\}}_{i\in I})$ in $Z^n(\cat{C})$ we define, by induction on $n\geq 0$,  a
  degeneracy map $d : N \to T$ and factorisations $A_i \to N \to X$ of each map $f_i$.
  For $n=0$, we set $d = \id$.  For $n > 0$, recall from Lemma~\ref{lemma:degeneracy-decomposition} that any degeneracy map $d: N \to T$ factors uniquely as a simple degeneracy map $d_S:N \to P$ followed by a parallel degeneracy map $d_P: P \to N$. Given the sink $\mathcal{S}$, we construct these maps $d_S$ and $d_P$ and the factorisations

\vspace{-20pt}
$$ A_i \to N\stackrel{d_S}{\to} P \stackrel{d_P}{\to} T$$ of $f_i: A_i \to T$   via the following scheme. 
  \begin{enumerate}
    \item Let $\reg{h}$ be a regular height of $T$. Consider the sink $$(T(\reg h),\{f_i(\reg h) : A_i(\reg{\regmap {f_i}(h)}) \to T(\reg{h})\})$$ in $Z^{n-1}(\cat{C})$ consisting of the component of the zigzag maps $f_i:A_i \to T$ at the regular height $\reg{h}$ (see Definition~\ref{definition:zigzag-map}).  Recursively apply relative normalisation to this sink to obtain an object $P(\reg{h})$ of $Z^{n-1}(\cat{C})$, a degeneracy map $P(\reg{h}) \to T(\reg{h})$, and for every $i \in I$ a factorisation as follows:
{\begin{equation}
        \label{diagram:normalisation-regular}
        \begin{tikzcd}
          A_i(\reg{\regmap{f_i}(h)}) \ar[dashed]{d} \ar{dr} \\[-7pt]
          P(\reg{h}) \ar[dashed]{r} & T(\reg{h})
        \end{tikzcd}
      \end{equation}}

    \item For every singular height $\sing{h}$ of $T$, consider the sink in $Z^{n-1}(\cat{C})$ which consists of the maps $f_i(\sing t):A_i(\sing{t}) \to T(\sing{h})$ for every $i \in I$ and every $t\in{(\singmap{f_i})}^{-1}(h)$, as well as the composite $P(\reg{h}) \to T(\reg{h}) \to T(\sing{h})$ and the composite $P(\reg{h} +1) \to T(\reg{h} +1) \to T(\sing{h})$. Recursively apply relative normalisation to this sink to obtain an object $P(\sing{h})$ in $Z^{n-1}(\cat C)$, a degeneracy map $P(\sing{h}) \to T(\sing{h})$, and the following factorisations, for every $i \in I$ and $t \in {(\singmap{f_i})}^{-1}(h)$:
{\begin{equation}
        \label{diagram:normalisation-singular-sink}
        \begin{tikzcd}
          A_i(\sing{t}) \ar[dashed]{d} \ar{dr}{f_i(\sing t)} \\[-7pt]
          P(\sing{h}) \ar[dashed]{r} & T(\sing{h})
        \end{tikzcd}
      \end{equation}      
      \begin{equation}
        \label{diagram:normalisation-singular-cospan}
        \begin{tikzcd}
          P(\reg{h}) \ar[dashed]{d} \ar{r}& T(\reg{h}) \ar{d} \\[-7pt]
          P(\sing{h}) \ar[dashed]{r} & T(\sing{h}) \\[-7pt]
          P(\reg{h + 1}) \ar[dashed]{u} \ar{r}& T(\reg{h + 1}) \ar{u}.
        \end{tikzcd}
      \end{equation}}
    \item The factorisations in (\ref{diagram:normalisation-singular-cospan}) assemble the objects $P(\reg{h})$ and $P(\sing{h})$ into a zigzag in $Z^{n-1}(\cat{C})$, and hence into an object $P$ of $Z^n(\cat{C})$. The degeneracy maps $P(\reg{h}) \to T(\reg{h})$ and $P(\sing{h}) \to T(\sing{h})$ assemble into a parallel 
      degeneracy map $d_P : P \to T$ in $Z^n(\cat{C})$. The factorisations of (\ref{diagram:normalisation-regular}) and (\ref{diagram:normalisation-singular-sink}) assemble into factorisations in~$Z^n(\cat{C})$:
{\[
        \begin{tikzcd}
          A_i \ar{d} \ar{dr}{f_i} \\[-7pt]
          P \ar{r} & T
        \end{tikzcd}
      \]}%
      Since the degeneracy map $P \to T$ is parallel, the maps $A_i \to P$ and $A_i \to T$ will have equal singular maps.

\vspace{4pt}
    \item 
    For those cospans $P(\reg{h}) \to P(\sing{h}) \leftarrow P(\reg{h}+1)$ with both legs given by isomorphisms, and for which the singular object $P(\sing{h})$ is not in the image of any of the $A_i \to P$, remove them from the zigzag $P \in Z^n(\cat C)$. This results in a smaller zigzag $N$, and a simple degeneracy map \mbox{$d_S:N \to P$} which re-inserts these trivial cospans. The maps $A_i \to P$ then canonically
      factor through this map, since by construction the removed heights are not in their image:
{\[
        \begin{tikzcd}
          A_i \ar{d} \ar{dr} \\[-5pt]
          N \ar{r} & P
        \end{tikzcd}
      \]}
    \item Define $d : N \to T$ to be the composite $d = d_P \circ d_S$.
  \end{enumerate}
\noindent
This concludes the description of the algorithm.

\begin{example}
\label{example:algorithm}
We illustrate the algorithm in Figure~\ref{fig:algorithm_illustration}, normalising a 1\-element sink $(T,\{f:A \to T\})$. On the left of the figure we show the structure of $A$, $T$ and $f$, while on the right of the figure we show the intermediate construction $P$, and the eventual normal form $N$.

\vspace{3pt}\noindent
-- \textit{Step 1.} Recursively apply relative normalisation to the 1\-element sink $(T(\reg 0),\{f(\reg 0):A(\reg 0) \to T(\reg 0)\})$ to obtain the factorization $$A(\reg 0) \to P(\reg 0) \stackrel {q(\reg 0)}\to T(\reg 0)$$and similarly for $f(\reg 1)$, $f(\reg 2)$, $f(\reg 3)$. This gives us the regular objects of $P$, and the regular slices $q(\reg i) : P(\reg i) \to T(\reg i)$.

\vspace{3pt}\noindent
-- \textit{Step 2.} Build the singular levels $P(\sing i)$, by recursively factorizing the sinks into $T(\sing i)$. For example, for $i=0$, we must factorize the following sink:
\begin{align*}
\Big( R(\sing 0), \big\{\,t_0 \circ q(\reg 0) &: P(\reg 0) \to T(\sing 0), t_0' \circ q(\reg 1) : P(\reg 1) \to T(\sing 0),
\\[-7pt]
&
\hspace{-40pt}
f(\sing 0): A(\sing 0) \to T(\sing 0), f(\sing 1) : A(\sing 1) \to T(\sing 0)
\,\,\big\} \Big)
\end{align*}
Factorizing this sink recursively yields the singular object $P(\sing 0)$ and the degeneracy maps $p_0: P(\reg 0) \to P(\sing 0)$ and \mbox{$p_0' : P(\reg 1) \to P(\sing 0)$}, as well as factorizing maps \mbox{$A(\sing 0) \to P(\sing 0)$} and $A(\sing 0) \to P(\sing 0)$.

\vspace{3pt}
\noindent
-- \textit{Step 3.} Assemble this data into the zigzag $P$, and the zigzag map $g:A \to N$, as shown in the figure.

\vspace{3pt}
\noindent
-- \textit{Step 4.} Inspect the maps $p_i, p_i'$ to find identity zigzags. We suppose for the sake of example that $p_1=\id$, $p_1' = \id$, $p_2 = \id$ and $p_2' = \id$. Since $T(\sing 1)$ is not in the image of $A \to T$, also $P(\sing 1)$ is not in the image of $A \to P$ (since those zigzag maps have equal singular maps), and we can therefore omit the entire $p_1,p_1'$ cospan, and we proceed to construct $N$ appropriately. Note that we retain the cospan $p_2,p_2'$  in $N$, even though the legs are identities, since $P(\sing 2)$ \textit{is} in the image of $A \to P$. The zigzag map $d_S : N \to P$ is then constructed as a simple degeneracy map, with face map omitting level 1.

\vspace{3pt}
\noindent
-- \textit{Step 5.} Produce the entire normalising degeneracy map as the composite $d_P \circ d_S : N \to T$.

\vspace{3pt}
\noindent
We are now done, and have factorized the original sink into the composite of a degeneracy map $d_p \circ d_s$, and a new sink $(N, \{ g: A \to N \})$.
\end{example}

\suckpara
\paragraph{Correctness.} We now show that Construction~\ref{constr:normalisation} correctly produces the relative normalisation of a sink.

\begin{proposition}\label{lemma:normalisation-correctness}
  Let  $\mathcal{S} = (T,\{ f_i : A_i \to T\}_i)$ be a sink in $Z^n(\cat{C})$. The map $d: N \to T$ constructed in Construction~\ref{constr:normalisation} is the relative normalisation of $T$ with respect to $\mathcal{S}$, i.e. the smallest element of $\Deg_{\mathcal{S}}(T)$. In particular, applied to the empty sink $\mathcal{S} = (T,\emptyset)$, the morphism $d:N \to T$ produced by Construction~\ref{constr:normalisation} is the normalisation of $T$.\end{proposition}

\begin{proof}
Recall that in Construction~\ref{constr:normalisation}, the degeneracy map \mbox{$d: N \to T$} is constructed as a composite $d_P \circ d_S$, where $d_P$ and $d_S$ are parallel and simple degeneracy maps respectively. We will prove the following three statements: 
  \begin{enumerate}
    \item $d_P$ is the initial parallel degeneracy map into $T$ through which the sink $\mathcal{S}$ factors.
    \item $d_S$ is the initial simple degeneracy map into $P$ such that the sink $\mathcal{S}$ factors through $d_P \circ d_S$.
    \item $d_P \circ d_S$ is the initial degeneracy map into $T$  through which $\mathcal{S}$ factors.  \end{enumerate}
Assume the inductive hypothesis that the claim holds in $Z^k(\cat{C})$ for $k < n$. We then proceed as follows.
  \begin{enumerate}
    \item
      Consider any parallel degeneracy map $d: P' \to P$, such that $\mathcal S$ factors through $d_P \circ d'$. Any regular slice map of $d_P \circ d'$ satisfies the
      factorisation condition (\ref{diagram:normalisation-regular}). Since Construction~\ref{constr:normalisation}
      has chosen the initial regular slice map for $d_P$ satisfying the conditions, the
      regular slices of $d'$ must be isomorphisms. But then the singular slice maps of $d_P \circ d'$ satisfy the factorisation
      conditions of (\ref{diagram:normalisation-singular-sink}) and (\ref{diagram:normalisation-singular-cospan}). 
      Similarly, it follows that the singular slices of $d'$
      must also be isomorphisms. So $d'$ is an isomorphism.
    \item
      The top-level degeneracy map $N \to P$ is chosen in Construction~\ref{constr:normalisation} to normalise
      as many trivial levels of $P$ as possible while retaining compatibility.

    \item Let $d' : N' \to T$ be any other degeneracy map via which $\mathcal{S}$ factors.
      By Lemma~\ref{lemma:degeneracy-decomposition}, $d'$ decomposes into a simple
      degeneracy map $d'_S : N' \to P'$ followed by a parallel degeneracy map $d'_P : P' \to T$.
      By part 1 there is a parallel degeneracy map $P \to P'$ which fits into this diagram:
      {\[
        \begin{tikzcd}
          P \ar{r}{d_P} \ar{d} & T \ar[equal]{d} \\[-5pt]
          P' \ar[swap]{r}{d_P'} & T
        \end{tikzcd}
      \]}%
      By Proposition~\ref{lemma:degeneracy-pullback} the pullback of $d_S' : N' \to P'$
      along $P \to P'$ exists and is a simple degeneracy map. But then by part 2
      there exists a map $N \to N' \times_{P'} P$ which makes the following
      diagram commute:
      {\[
        \begin{tikzcd}
          N \ar{d} \ar{dr}{d_S}\\[-7pt]
          N' \times_{P'} P \ar{r} \ar{d} & P \ar{r}{d_P} \ar{d} & T \ar[equal]{d} \\[-8pt]
          N' \ar[swap]{r}{d_S'} & P' \ar[swap]{r}{d_P'} & T
        \end{tikzcd}
      \]}%
      So $d_P \circ d_S$ represents a smaller subobject of $T$. The claim
      follows since $d'$ was chosen arbitrarily among the compatible degeneracy maps.\qedhere
  \end{enumerate}
\end{proof}

\suckpara
\paragraph{Reflective Localisation}
This relative sink normalisation may be considered a special case of the following general machinery. Given an object $T$ in a category $\cat{A}$  and a class of morphisms $\mathfrak{D}$ in $\cat{A}$, let $\cat{A}/T$ denote the over-category, and consider the inclusion of the full subcategory $\cat{A}/^{\mathfrak{D}}T \hookrightarrow \cat{A}/T$ of those $d:A \to T$ which are in $\mathfrak{D}$. This inclusion has a left-adjoint $L: \cat{A}/T \to \cat{A}/^{\mathfrak{D}}T$ if and only if, for every morphism $f:A \to T$ in $\cat{A}$, the evident category of factorisations of $f$ into a morphism in $\cat{A}$ followed by a morphism in $\mathfrak{D}$ has an initial object. The image $L(f) \in \cat{A}/^{\mathfrak{D}}{T}$ is the $\mathfrak{D}$-morphism part of this initial factorisation of $f$. 

An analogous observation applies to the full inclusion $\cat{A}{/^\mathfrak{D}T} \hookrightarrow \mathrm{Sink}_{\cat{A}}(T)$, where $\mathrm{Sink}_{\cat{A}}(T)$ is an appropriate category of sinks $\mathcal{S}=(T,\{f_i:A_i \to T\})$ in $\cat{A}$ into $T$. This inclusion has a left adjoint if and only if, for every sink $\mathcal{S}$, the associated category of factorisations of $\mathcal{S}$ into a sink $(N,\{g_i: A_i \to N\})$ followed by a morphism $N \to T$ in $\mathfrak{D}$ has an initial object.

Applied to the situation where $\cat{A}= Z^n(\cat{C})$ and $\mathfrak{D}$ is the class of degeneracy maps,  Proposition~\ref{lemma:initialdegeneratesink} may therefore be understood as asserting that for every object $T\in Z^n (\cat{C})$, the inclusion

\vspace{-20pt}
$$R:Z^n(\cat{C})/^{\mathrm{deg}}{T} \to \mathrm{Sink}_{Z^n(\cat{C})}(T)$$ has a left adjoint $L$. The relative normalisation of a sink \mbox{$(T,\{f_i:A_i \to T\})$} is then constructed as its image under $L$.

Following standard terminology~\cite[\S~IV.3]{MacLane}, this says that $Z^n(\cat{C}){/^{\mathrm{deg}}T}$ is a \textit{reflective subcategory} of  $\mathrm{Sink}_{Z^n(\cat{C})}(T)$, and relative normalisation is the corresponding \textit{reflective localisation} functor $\mathrm{Sink}_{Z^n(\cat{C})}(T) \to Z^n(\cat{C})/^{\mathrm{deg}}{T}$. 
\\ 
\sucksec
\section{Globularity}\label{sec:globularity}

Associative $n$-categories form a globular theory of higher categories, meaning that for any diagram, the boundary of the source matches the boundary of the target. This is enforced in the proof assistant by requiring that diagrams have a globularity property, meaning intuitively that regular slices have to act trivially. We define this formally as follows.

\begin{definition}
  In $Z^n(\cat C)$, a map $f$ is a \textit{globular map} if $n = 0$, or both the following properties hold:
  \begin{enumerate}
    \item all regular slice maps of $f$ are isomorphisms;
    \item all singular slice maps of $f$ are globular in $Z^{n - 1}(\cat{C})$.
  \end{enumerate}
  An object of $Z^n(\cat{C})$ is a \textit{globular object} if $n = 0$, or it is a zigzag of
  globular objects and globular maps in $Z^{n - 1}(\cat{C})$.
\end{definition}

\noindent
To be valid in the proof assistant, a diagram must be globular, and its normalisation must also be globular. It is therefore a requirement that normalization preserves globularity, and we verify this here.

The core of the argument is that the normalisation algorithm maintains the
invariant that all maps in the sinks of the recursive applications
already normalise the regular levels, so the factorisations can be globular.
We define this property formally as follows.

\begin{definition}
  In $Z^n(\cat C)$, a map is \textit{regularly normalising} if $n = 0$, or both the following properties hold:

\vspace{2pt}
\noindent
\begin{minipage}{\linewidth}
  \begin{enumerate}
    \item all regular slice maps are normalising;
    \nopagebreak
    \item all singular slice maps are regularly normalising.
  \end{enumerate}
  \end{minipage}
\end{definition}

\begin{lemma}\label{lemma:normalisation-invariant}
  Let $d : N \to X$ be the relative normalisation of a globular object $X \in
  Z^n(\cat{C})$ with respect to a sink $(X,{\{ f_i: A_i \to X \}}_i)$ of regularly normalising maps. Then $N$
  is a globular object, the factorisations of the maps in the sink are globular
  maps and $d$ is regularly normalising.
\end{lemma}
\begin{proof}
  By Proposition~\ref{lemma:normalisation-correctness} the normalisation
  algorithm correctly computes the relative normalisation. 
  Since the maps $f_i$ are regularly normalising, the regular slices of $A_i$
  are already normalised and so the algorithm fills the diagram (\ref{diagram:normalisation-regular})
  as follows:
  {\begin{equation*}
    \begin{tikzcd}
      A(\reg{ \regmap f (h)}) \ar[swap]{d}{\id} \ar{dr}{f(\reg{i})} \\[-3pt]
      P(\reg{h}) \ar{r} & X(\reg{h})
    \end{tikzcd}
  \end{equation*}}%
  In particular, the regular slices of the computed factorisations $A_i \to P$
  are identities and the regular slices of the degeneracy map $d_P$ are
  normalising. 

  Since $X$ is globular so are its singular slices. Since the sink consists of regularly normalising maps, the solid maps arising from
  (\ref{diagram:normalisation-singular-sink}) are regularly
  normalising maps into globular objects. The maps $P(\reg{h}) \to X(\sing{h})$
  and $P(\reg{h + 1}) \to X(\sing{h})$ in (\ref{diagram:normalisation-singular-cospan})
  are composites of a normalising map followed by a globular one, so they are
certainly  regularly normalising. Therefore the sinks formed in (\ref{diagram:normalisation-singular-sink})
  and (\ref{diagram:normalisation-singular-cospan}) satisfy the conditions of
  this Lemma. By induction the singular slices of $d_P$ are regularly normalising,
  the singular slices of factorisations $A_i \to P$ are globular,
  and the cospans $P(\reg{h}) \to P(\sing{h}) \leftarrow P(\reg{h + 1})$ are globular
  maps between globular objects.

  By the observations above, the parallel degeneracy map $d_P$ is regularly
  normalising, $P$ is a globular object and the factorisations $A_i \to P$
  are globular maps. These properties are preserved by the final step
  which precomposes $d_P$ by the simple degeneracy map $d_S$.
\end{proof}

\begin{proposition}\label{prop:normalisation-globular}
  The normalisation of a globular object is globular.
\end{proposition}
\begin{proof}
  The normalisation of some globular object $X$ is the relative normalisation of $X$
  with respect to the empty sink. Thus the result follows by Lemma~\ref{lemma:normalisation-invariant}.
\end{proof}

The invariants of Lemma~\ref{lemma:normalisation-invariant} can also be of help
in the implementation of the normalisation algorithm. The regular slices of the
degeneracy maps are determined by their targets and thus do not need to be
represented explicitly. All diagrams are globular, as well as all the factorisation maps, allowing them to be represented by simpler data structures for which globularity is hard-coded. The non-globular sink maps can be represented as formal composites of a degeneracy map followed by a globular map.

\section{Examples}
\label{sec:implementation}

In this section we sketch the type checking scheme, and show some worked examples of interest in higher category theory, the \textit{Eckmann-Hilton Move} and the \textit{Syllepsis}.

\paragraph{Type Checking.}
\def\SV{\mathsf{SV}}

We first give an informal overview of type checking, focusing on its relevance for normalisation.

For an $n$\-diagram $D$ given as an object of $Z^n(\N)$, we define its \emph{singular content} as a 1\-element set if \mbox{$n=0$}, or else by recursion as the disjoint union of the singular content of its singular objects. For example, the 2\-diagram of Figure~\ref{fig:zigzagmotivation} has singular content of cardinality 10. We then break $D$ into a number of \textit{pieces}, one for each element of singular content, by taking the preimages of the elements of singular content under the singular map structures defining $D$.

The type checking procedure works with respect to a \emph{signature} of allowed algebraic generators. Given a globular $n$\-diagram, we normalize each piece, and then check if the resulting $n$\-diagram is an element of the signature. If this is the case for all pieces, the diagram is declared valid.

\suckpara
\paragraph{Examples.}

Here we illustrate the type checking procedure for two examples. Although we label points of diagrams in this section with generator names, for the purpose of normalisation we implicitly use the generator dimensions to obtain an $\N$-labelling, as on the right of Figure~\ref{fig:zigzagmotivation}.

Each example is accompanied by a hyperlink to the type-checked   formalisation in the proof assistant, which will display a 3d model. Left-click and drag to rotate; right-click and drag to pan; use the mouse wheel to zoom. A video is also provided for each example, showing how it is constructed.

\suckpara
\begin{example} \ 

\vspace{2pt}

\noindent \textit{Video:} \url{https://youtu.be/lk-71EsZqaE}

\vspace{2pt}
\noindent
The \textit{Eckmann-Hilton Move} is a 3-morphism in a finitely presented 3-category, generated by a single 0-cell  $\bullet$, and 2-cells \mbox{$x,y:\id(\bullet) \to \id(\bullet)$}. The signature therefore comprises the following nontrivial diagrams:
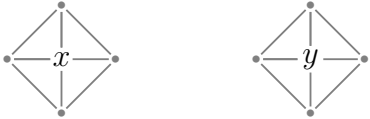
\def\extrascale{1.2}
$$
\begin{tikzpicture}[xscale=1, scale=.6, scale=\extrascale]
\node [v] (1) at (0,-0) {$\bu$};
\node [v] (7) at (0,2) {$\bu$};
\node [v] (2) at (-1,1) {$\bu$};
\node [v] (3) at (0,1) {$x$};
\node [v] (6) at (1,1) {$\bu$};
\begin{pgfonlayer}{background}
\ff 1 3 \ee 3 7 \ff 3 7 \ff 2 3 \ff 2 6 \ff 2 7 \ff 6 7 \ff 1 2 \ff 1 6
\end{pgfonlayer}
\end{tikzpicture}
\hspace{1.7cm}
\begin{tikzpicture}[xscale=1, scale=.6, scale=\extrascale]
\node [v] (1) at (0,-0) {$\bu$};
\node [v] (7) at (0,2) {$$\bu$$};
\node [v] (2) at (-1,1) {$\bu$};
\node [v] (3) at (0,1) {$y$};
\node [v] (6) at (1,1) {$\bu$};
\begin{pgfonlayer}{background}
\ff 1 3 \ee 3 7 \ff 3 7 \ff 2 3 \ff 2 6 \ff 2 7 \ff 6 7 \ff 1 2 \ff 1 6
\end{pgfonlayer}
\end{tikzpicture}
$$
The Eckmann-Hilton Move itself is represented by the following 3-diagram, interpreted as $x$ ``braiding'' around $y$:
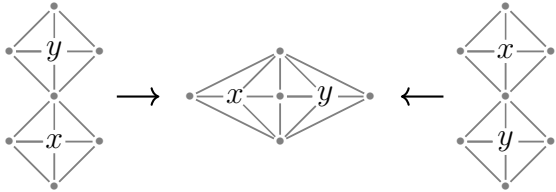
$$
\begin{tikzpicture}[xscale=1]
\node (a) at (-1,0) {\begin{tikzpicture}[xscale=1, scale=.5, scale=\extrascale]
\node [v] (1) at (0,-0) {$\bu$};
\node [v] (7) at (0,2) {$$\bu$$};
\node [v] (2) at (-1,1) {$\bu$};
\node [v] (3) at (0,1) {$x$};
\node [v] (6) at (1,1) {$\bu$};
\node [v] (7') at (0,4) {$$\bu$$};
\node [v] (2') at (-1,3) {$\bu$};
\node [v] (3') at (0,3) {$y$};
\node [v] (6') at (1,3) {$\bu$};
\begin{pgfonlayer}{background}
\ff 1 3 \ee 3 7 \ff 2 3 \ff 2 6 \ff 2 7 \ff 6 7 \ff 1 2 \ff 1 6
\ff {7} {3'} \ee {3'} {7'} \ff {2'} {3'} \ff {2'} {6'} \ff {2'} {7'} \ff {6'} {7'} \ff {7} {2'} \ff {7} {6'}
\end{pgfonlayer}
\end{tikzpicture}};

\node (b) at (2,0) {\begin{tikzpicture}[xscale=1, scale=.5, scale=\extrascale]
\node [v] (1) at (-1,-0) {$\bu$};
\node [v] (7) at (-1,2) {$$\bu$$};
\node [v] (2) at (-1,1) {$\bu$};
\node [v] (3) at (0,1) {$y$};
\node [v] (6) at (1,1) {$\bu$};
\node [v] (2') at (-3,1) {$\bu$};
\node [v] (3') at (-2,1) {$x$};
\begin{pgfonlayer}{background}
\ff 1 3 \ee 3 7 \ff 2 3 \ff 2 6 \ff 2 7 \ff 6 7 \ff 1 2 \ff 1 6
\ff {7} {3'} \ee {3'} {1} \ff {2'} {3'} \ff {2'} {1} \ff {7} {2'} \ff {3'} 2
\end{pgfonlayer}
\end{tikzpicture}};

\node (c) at (5,0) {\begin{tikzpicture}[xscale=1, scale=.5, scale=\extrascale]
\node [v] (1) at (0,-0) {$\bu$};
\node [v] (7) at (0,2) {$$\bu$$};
\node [v] (2) at (-1,1) {$\bu$};
\node [v] (3) at (0,1) {$y$};
\node [v] (6) at (1,1) {$\bu$};
\node [v] (7') at (0,4) {$$\bu$$};
\node [v] (2') at (-1,3) {$\bu$};
\node [v] (3') at (0,3) {$x$};
\node [v] (6') at (1,3) {$\bu$};
\begin{pgfonlayer}{background}
\ff 1 3 \ee 3 7 \ff 2 3 \ff 2 6 \ff 2 7 \ff 6 7 \ff 1 2 \ff 1 6
\ff {7} {3'} \ee {3'} {7'} \ff {2'} {3'} \ff {2'} {6'} \ff {2'} {7'} \ff {6'} {7'} \ff {7} {2'} \ff {7} {6'}
\end{pgfonlayer}
\end{tikzpicture}};

\draw [->, thick, shorten >=5pt] (a) to (b);
\draw [->, thick, shorten >=5pt] (c) to (b);

\end{tikzpicture}
$$
This has singular content $\{x,y\}$. The piece containing singular content $x$ is the following 3\-diagram, which we name~$D$:
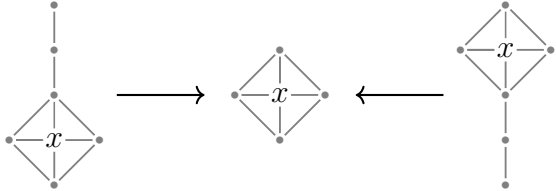
$$
\begin{tikzpicture}[xscale=1]
\node (a) at (-1,0) {\begin{tikzpicture}[xscale=1, scale=.5, scale=\extrascale]
\node [v] (1) at (0,-0) {$\bu$};
\node [v] (7) at (0,2) {$$\bu$$};
\node [v] (2) at (-1,1) {$\bu$};
\node [v] (3) at (0,1) {$x$};
\node [v] (6) at (1,1) {$\bu$};
\node [v] (7') at (0,4) {$$\bu$$};
\node [v] (2') at (0,3) {$\bu$};
\begin{pgfonlayer}{background}
\ff 1 3 \ee 3 7 \ff 2 3 \ff 2 6 \ff 2 7 \ff 6 7 \ff 1 2 \ff 1 6
\ff {2'} {7'} \ff {7} {2'}
\end{pgfonlayer}
\end{tikzpicture}};

\node (b) at (2,0) {\begin{tikzpicture}[xscale=1, scale=.5, scale=\extrascale]
\node [v] (1) at (0,-0) {$\bu$};
\node [v] (7) at (0,2) {$\bu$};
\node [v] (2) at (-1,1) {$\bu$};
\node [v] (3) at (0,1) {$x$};
\node [v] (6) at (1,1) {$\bu$};
\begin{pgfonlayer}{background}
\ff 1 3 \ee 3 7 \ff 3 7 \ff 2 3 \ff 2 6 \ff 2 7 \ff 6 7 \ff 1 2 \ff 1 6
\end{pgfonlayer}
\end{tikzpicture}};

\node (c) at (5,0) {\begin{tikzpicture}[xscale=1, scale=.5, scale=\extrascale]
\node [v] (1) at (0,-0) {$\bu$};
\node [v] (7) at (0,2) {$\bu$};
\node [v] (2) at (0,1) {$\bu$};
\node [v] (7') at (0,4) {$\bu$};
\node [v] (2') at (-1,3) {$\bu$};
\node [v] (3') at (0,3) {$x$};
\node [v] (6') at (1,3) {$\bu$};
\begin{pgfonlayer}{background}
\ff 2 7 \ff 1 2
\ff {7} {3'} \ee {3'} {7'} \ff {2'} {3'} \ff {2'} {6'} \ff {2'} {7'} \ff {6'} {7'} \ff {7} {2'} \ff {7} {6'}
\end{pgfonlayer}
\end{tikzpicture}};

\draw [->, thick, shorten >=5pt] (a) to (b);
\draw [->, thick, shorten >=5pt] (c) to (b);

\end{tikzpicture}
$$
To normalise this 3-diagram piece we apply our normalisation algorithm, as presented in Construction~\ref{constr:normalisation}. Step~1 invokes recursive calls which normalise the left and right boundaries of $D$, with the following results:
$$
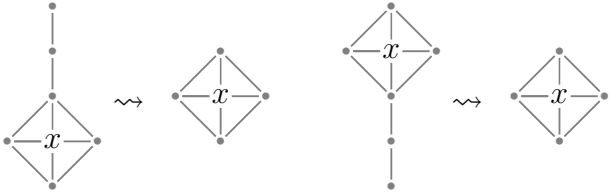
\begin{aligned}
\begin{tikzpicture}[xscale=1, scale=.5, scale=\extrascale]
\node [v] (1) at (0,-0) {$\bu$};
\node [v] (7) at (0,2) {$$\bu$$};
\node [v] (2) at (-1,1) {$\bu$};
\node [v] (3) at (0,1) {$x$};
\node [v] (6) at (1,1) {$\bu$};
\node [v] (7') at (0,4) {$$\bu$$};
\node [v] (2') at (0,3) {$\bu$};
\begin{pgfonlayer}{background}
\ff 1 3 \ee 3 7 \ff 2 3 \ff 2 6 \ff 2 7 \ff 6 7 \ff 1 2 \ff 1 6
\ff {2'} {7'} \ff {7} {2'}
\end{pgfonlayer}
\end{tikzpicture}
\end{aligned}
\leadsto
\,\,\,
\begin{aligned}
\begin{tikzpicture}[xscale=1, scale=.5, scale=\extrascale]
\node [v] (1) at (0,-0) {$\bu$};
\node [v] (7) at (0,2) {$$\bu$$};
\node [v] (2) at (-1,1) {$\bu$};
\node [v] (3) at (0,1) {$x$};
\node [v] (6) at (1,1) {$\bu$};
\begin{pgfonlayer}{background}
\ff 1 3 \ee 3 7 \ff 2 3 \ff 2 6 \ff 2 7 \ff 6 7 \ff 1 2 \ff 1 6
\end{pgfonlayer}
\end{tikzpicture}
\end{aligned}
\hspace{.9cm}
\begin{aligned}
\begin{tikzpicture}[xscale=1, scale=.5, scale=\extrascale]
\node [v] (1) at (0,-0) {$\bu$};
\node [v] (7) at (0,2) {$\bu$};
\node [v] (2) at (0,1) {$\bu$};
\node [v] (7') at (0,4) {$\bu$};
\node [v] (2') at (-1,3) {$\bu$};
\node [v] (3') at (0,3) {$x$};
\node [v] (6') at (1,3) {$\bu$};
\begin{pgfonlayer}{background}
\ff 2 7 \ff 1 2
\ff {7} {3'} \ee {3'} {7'} \ff {2'} {3'} \ff {2'} {6'} \ff {2'} {7'} \ff {6'} {7'} \ff {7} {2'} \ff {7} {6'}
\end{pgfonlayer}
\end{tikzpicture}\end{aligned}
\leadsto
\,\,\,
\begin{aligned}
\begin{tikzpicture}[xscale=1, scale=.5, scale=\extrascale]
\node [v] (1) at (0,-0) {$\bu$};
\node [v] (7) at (0,2) {$$\bu$$};
\node [v] (2) at (-1,1) {$\bu$};
\node [v] (3) at (0,1) {$x$};
\node [v] (6) at (1,1) {$\bu$};
\begin{pgfonlayer}{background}
\ff 1 3 \ee 3 7 \ff 2 3 \ff 2 6 \ff 2 7 \ff 6 7 \ff 1 2 \ff 1 6
\end{pgfonlayer}
\end{tikzpicture}
\end{aligned}
$$
In Steps 2 and  3, we use these results to obtain the intermediate normalisation zigzag  $P$, a 3\-diagram of length 1:
$$
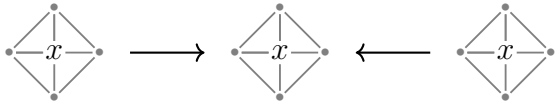
\begin{tikzpicture}[xscale=1]
\node (a) at (-1,0) {\begin{tikzpicture}[xscale=1, scale=.5, scale=\extrascale]
\node [v] (1) at (0,-0) {$\bu$};
\node [v] (7) at (0,2) {$\bu$};
\node [v] (2) at (-1,1) {$\bu$};
\node [v] (3) at (0,1) {$x$};
\node [v] (6) at (1,1) {$\bu$};
\begin{pgfonlayer}{background}
\ff 1 3 \ee 3 7 \ff 3 7 \ff 2 3 \ff 2 6 \ff 2 7 \ff 6 7 \ff 1 2 \ff 1 6
\end{pgfonlayer}
\end{tikzpicture}};

\node (b) at (2,0) {\begin{tikzpicture}[xscale=1, scale=.5, scale=\extrascale]
\node [v] (1) at (0,-0) {$\bu$};
\node [v] (7) at (0,2) {$\bu$};
\node [v] (2) at (-1,1) {$\bu$};
\node [v] (3) at (0,1) {$x$};
\node [v] (6) at (1,1) {$\bu$};
\begin{pgfonlayer}{background}
\ff 1 3 \ee 3 7 \ff 3 7 \ff 2 3 \ff 2 6 \ff 2 7 \ff 6 7 \ff 1 2 \ff 1 6
\end{pgfonlayer}
\end{tikzpicture}};

\node (c) at (5,0) {\begin{tikzpicture}[xscale=1, scale=.5, scale=\extrascale]
\node [v] (1) at (0,-0) {$\bu$};
\node [v] (7) at (0,2) {$\bu$};
\node [v] (2) at (-1,1) {$\bu$};
\node [v] (3) at (0,1) {$x$};
\node [v] (6) at (1,1) {$\bu$};
\begin{pgfonlayer}{background}
\ff 1 3 \ee 3 7 \ff 3 7 \ff 2 3 \ff 2 6 \ff 2 7 \ff 6 7 \ff 1 2 \ff 1 6
\end{pgfonlayer}
\end{tikzpicture}};

\draw [->, thick, shorten >=5pt, shorten <=5pt] (a) to (b);
\draw [->, thick, shorten >=5pt, shorten <=5pt] (c) to (b);

\end{tikzpicture}
$$
We note that this is an identity cospan, and so in Step 4 of the algorithm we omit this cospan when we form $N$: 
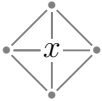
$$
\begin{tikzpicture}[xscale=1]
\node (c) at (5,0) {\begin{tikzpicture}[xscale=1, scale=.5, scale=\extrascale]
\node [v] (1) at (0,-0) {$\bu$};
\node [v] (7) at (0,2) {$\bu$};
\node [v] (2) at (-1,1) {$\bu$};
\node [v] (3) at (0,1) {$x$};
\node [v] (6) at (1,1) {$\bu$};
\begin{pgfonlayer}{background}
\ff 1 3 \ee 3 7 \ff 3 7 \ff 2 3 \ff 2 6 \ff 2 7 \ff 6 7 \ff 1 2 \ff 1 6
\end{pgfonlayer}
\end{tikzpicture}};
\end{tikzpicture}
$$
This is the normal form of our original piece $D$. This is an element of  our signature, hence the piece $D$ is validated by the type checker. The piece corresponding to $y$ is also valid, and so the entire Eckmann--Hilton 3-diagram type checks.
\end{example}

\suckpara
\suckpara
\begin{example} \ 


\vspace{2pt}
\noindent \textit{Video:} \url{http://youtu.be/76UJgg-ibO8}

\vspace{2pt}
\noindent
The \textit{Syllepsis} is a 5\-morphism in a finitely presented 5-category, generated by a single 0-cell $\bullet$, and two 3-cells with types  \mbox{$x,y:\id(\id(\bullet)) \to \id(\id(\bullet))$}. The signature therefore contains these 3-diagrams, which we draw in a quasi-3d style:
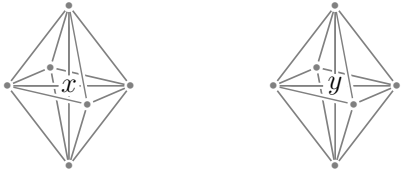
\def\extrascale{1.3}
$$
\tikz{\node [scale=.9] at (0,0) {\begin{tikzpicture}[xscale=1, scale=.7, scale=\extrascale]
\node [v] (1) at (0,-.3) {$\bu$};
\node [v] (7) at (0,2.3) {$$\bu$$};
\node [v] (2) at (-1,1) {$\bu$};
\node [v] (3) at (0,1) {$x$};
\node [v] (4) at (-\xo,1+\yo) {$\bu$};
\node [v] (5) at (\xo,1-\yo) {$\bu$};
\node [v] (6) at (1,1) {$\bu$};
\begin{pgfonlayer}{background}
\ff 1 2 \ff 1 3 \ff 1 4 \ff 1 5 \ff 1 6 \ee 2 3 \ee 2 5 \ff 2 4 \ff 2 3 \ff 5 6 \ff 3 6 \ff 4 6 \ff 2 7 \ee 3 7 \ee 5 7 \ff 4 7 \ff 3 7 \ff 6 7
\end{pgfonlayer}
\begin{pgfonlayer}{background}
\end{pgfonlayer}
\end{tikzpicture}}}
\hspace{1.5cm}
\tikz{\node [scale=.9] at (0,0) {\begin{tikzpicture}[xscale=1, scale=.7, scale=\extrascale]
\node [v] (1) at (0,-.3) {$\bu$};
\node [v] (7) at (0,2.3) {$$\bu$$};
\node [v] (2) at (-1,1) {$\bu$};
\node [v] (3) at (0,1) {$y$};
\node [v] (4) at (-\xo,1+\yo) {$\bu$};
\node [v] (5) at (\xo,1-\yo) {$\bu$};
\node [v] (6) at (1,1) {$\bu$};
\begin{pgfonlayer}{background}
\ff 1 2 \ff 1 3 \ff 1 4 \ff 1 5 \ff 1 6 \ee 2 3 \ee 2 5 \ff 2 4 \ff 2 3 \ff 5 6 \ff 3 6 \ff 4 6 \ff 2 7 \ee 3 7 \ee 5 7 \ff 4 7 \ff 3 7 \ff 6 7
\end{pgfonlayer}
\begin{pgfonlayer}{background}
\end{pgfonlayer}
\end{tikzpicture}}}
$$
We depict the Syllepsis 5-diagram in Figure~\ref{fig:syllepsis}. Intuitively, it represents the equivalence between the braid and its inverse when immersed in 4\-dimensional space. In the live proof, use the ``Slice'' control on the right to navigate through this equivalence. It has singular content $\{x,y\}$, and we extract the piece containing singular content $x$, depicting it in Figure~\ref{fig:syllepsis_x}. Applying our normalisation algorithm, following a long sequence of recursive calls, we obtain the normal form:
$$
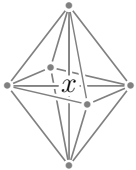
\tikz{\node [scale=.9] at (0,0) {\begin{tikzpicture}[xscale=1, scale=.7, scale=\extrascale]
\node [v] (1) at (0,-.3) {$\bu$};
\node [v] (7) at (0,2.3) {$$\bu$$};
\node [v] (2) at (-1,1) {$\bu$};
\node [v] (3) at (0,1) {$x$};
\node [v] (4) at (-\xo,1+\yo) {$\bu$};
\node [v] (5) at (\xo,1-\yo) {$\bu$};
\node [v] (6) at (1,1) {$\bu$};
\begin{pgfonlayer}{background}
\ff 1 2 \ff 1 3 \ff 1 4 \ff 1 5 \ff 1 6 \ee 2 3 \ee 2 5 \ff 2 4 \ff 2 3 \ff 5 6 \ff 3 6 \ff 4 6 \ff 2 7 \ee 3 7 \ee 5 7 \ff 4 7 \ff 3 7 \ff 6 7
\end{pgfonlayer}
\begin{pgfonlayer}{background}
\end{pgfonlayer}
\end{tikzpicture}}}
$$
Since this is an element of our signature, we determine that the piece is valid. Similarly, the piece corresponding to singular content $y$ is valid, and hence the entire Syllepsis 5\-diagram type checks.
\end{example}


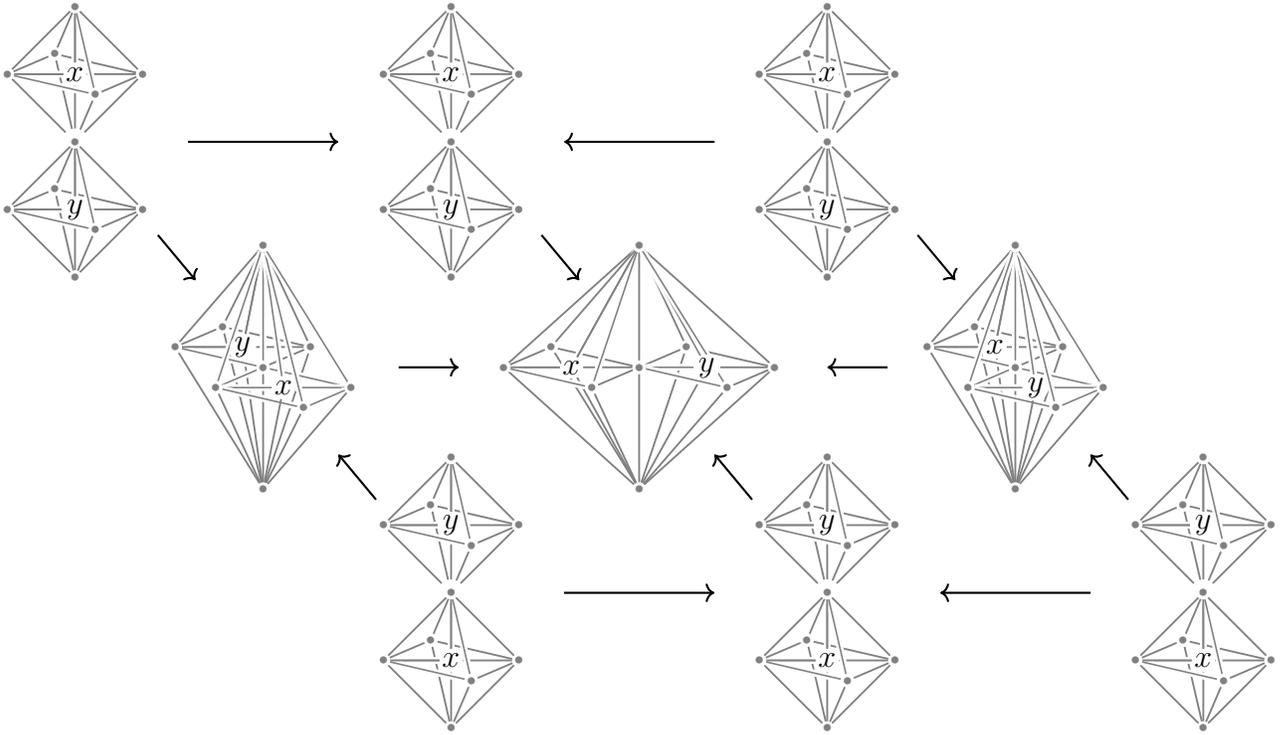
\begin{figure*}[h!]
$
\begin{tikzpicture}[yscale=1.0]

\def\microscale{.9}

\node (a) at (5,0) {$\begin{tikzpicture}[scale=\microscale]
\node [v] (1) at (0,0) {$\bu$};
\node [v] (2) at (-1,1) {$\bu$};
\node [v] (3) at (0,1) {$x$};
\node [v] (4) at (-\xo,1+\yo) {$\bu$};
\node [v] (5) at (\xo,1-\yo) {$\bu$};
\node [v] (6) at (1,1) {$\bu$};
\node [v] (7) at (0,2) {$\bu$};
\begin{pgfonlayer}{background}
\ff 1 2 \ff 1 3 \ff 1 4 \ff 1 5 \ff 1 6 \ee 2 3 \ee 2 5 \ff 2 4 \ff 2 3 \ff 5 6 \ff 3 6 \ff 4 6 \ff 2 7 \ee 3 7 \ee 5 7 \ff 4 7 \ff 3 7 \ff 6 7
\end{pgfonlayer}
\begin{scope}[yshift=2cm]
\node [v, white, opacity=0] (1) at (0,0) {7};
\node [v] (2) at (-1,1) {\bu};
\node [v] (3) at (0,1) {$y$};
\node [v] (4) at (-\xo,1+\yo) {\bu};
\node [v] (5) at (\xo,1-\yo) {$\bu$};
\node [v] (6) at (1,1) {\bu};
\node [v] (7) at (0,2) {\bu};
\begin{pgfonlayer}{background}
\ff 1 2 \ff 1 3 \ff 1 4 \ff 1 5 \ff 1 6 \ee 2 3 \ee 2 5 \ff 2 4 \ff 2 3 \ff 5 6 \ff 3 6 \ff 4 6 \ff 2 7 \ee 3 7 \ee 5 7 \ff 4 7 \ff 3 7 \ff 6 7
\end{pgfonlayer}
\end{scope}
\end{tikzpicture}
$};
\node (b) at (2.5,3) {$\def\xo{.3}
\begin{tikzpicture}[scale=\microscale]
\node [v] (1) at (0,-.8) {$\bu$};
\node [v] (2) at (-1-\xo,1+\yo) {$\bu$};
\node [v] (3) at (-\xo,1+\yo) {$y$};
\node [v] (4) at (-2*\xo,1+2*\yo) {$\bu$};
\node [v] (5) at (0,1) {$\bu$};
\node [v] (6) at (1-\xo,1+\yo) {$\bu$};
\node [v] (7) at (0,2.8) {$\bu$};
\begin{pgfonlayer}{background}
\ff 1 2 \ff 1 3 \ff 1 4 \ff 1 5 \ff 1 6 \ee 2 3 \ee 2 5 \ff 2 4 \ff 2 3 \ff 5 6 \ff 3 6 \ff 4 6 \ff 2 7 \ee 3 7 \ee 5 7 \ff 4 7 \ff 3 7 \ff 6 7
\end{pgfonlayer}
\node [v] (11) at (\xo-1,1-\yo) {$\bu$};
\node [v] (8) at (\xo,1-\yo) {$x$};
\node [v] (9) at (2*\xo,1-2*\yo) {$\bu$};
\node [v] (10) at (1+\xo,1-\yo) {$\bu$};
\begin{pgfonlayer}{background}
\ee 5 {11} \ee 5 {10} \ee 8 {10} \ee 9 {10} \ee 8 {11} \ee 9 {11} \ff 1 {11} \ff 1 9 \ff 1 8 \ff 1 {10} \ee {11} 7 \ee 8 7 \ee 9 7 \ee {10} 7
\end{pgfonlayer}
\end{tikzpicture}
$};
\node (c) at (0,6) {$\begin{tikzpicture}[scale=\microscale]
\node [v] (1) at (0,0) {$\bu$};
\node [v] (2) at (-1,1) {\bu};
\node [v] (3) at (0,1) {$y$};
\node [v] (4) at (-\xo,1+\yo) {$\bu$};
\node [v] (5) at (\xo,1-\yo) {$\bu$};
\node [v] (6) at (1,1) {$\bu$};
\node [v] (7) at (0,2) {$\bu$};
\begin{pgfonlayer}{background}
\ff 1 2 \ff 1 3 \ff 1 4 \ff 1 5 \ff 1 6 \ee 2 3 \ee 2 5 \ff 2 4 \ff 2 3 \ff 5 6 \ff 3 6 \ff 4 6 \ff 2 7 \ee 3 7 \ee 5 7 \ff 4 7 \ff 3 7 \ff 6 7
\end{pgfonlayer}
\begin{scope}[yshift=2cm]
\node [v, white, opacity=0] (1) at (0,0) {7};
\node [v] (2) at (-1,1) {$\bu$};
\node [v] (3) at (0,1) {$x$};
\node [v] (4) at (-\xo,1+\yo) {$\bu$};
\node [v] (5) at (\xo,1-\yo) {$\bu$};
\node [v] (6) at (1,1) {$\bu$};
\node [v] (7) at (0,2) {$\bu$};
\begin{pgfonlayer}{background}
\ff 1 2 \ff 1 3 \ff 1 4 \ff 1 5 \ff 1 6 \ee 2 3 \ee 2 5 \ff 2 4 \ff 2 3 \ff 5 6 \ff 3 6 \ff 4 6 \ff 2 7 \ee 3 7 \ee 5 7 \ff 4 7 \ff 3 7 \ff 6 7
\end{pgfonlayer}
\end{scope}
\end{tikzpicture}
$};


\node (a) at (10,0) {$\begin{tikzpicture}[scale=\microscale]
\node [v] (1) at (0,0) {$\bu$};
\node [v] (2) at (-1,1) {$\bu$};
\node [v] (3) at (0,1) {$x$};
\node [v] (4) at (-\xo,1+\yo) {$\bu$};
\node [v] (5) at (\xo,1-\yo) {$\bu$};
\node [v] (6) at (1,1) {$\bu$};
\node [v] (7) at (0,2) {$\bu$};
\begin{pgfonlayer}{background}
\ff 1 2 \ff 1 3 \ff 1 4 \ff 1 5 \ff 1 6 \ee 2 3 \ee 2 5 \ff 2 4 \ff 2 3 \ff 5 6 \ff 3 6 \ff 4 6 \ff 2 7 \ee 3 7 \ee 5 7 \ff 4 7 \ff 3 7 \ff 6 7
\end{pgfonlayer}
\begin{scope}[yshift=2cm]
\node [v, white, opacity=0] (1) at (0,0) {7};
\node [v] (2) at (-1,1) {\bu};
\node [v] (3) at (0,1) {$y$};
\node [v] (4) at (-\xo,1+\yo) {\bu};
\node [v] (5) at (\xo,1-\yo) {$\bu$};
\node [v] (6) at (1,1) {\bu};
\node [v] (7) at (0,2) {\bu};
\begin{pgfonlayer}{background}
\ff 1 2 \ff 1 3 \ff 1 4 \ff 1 5 \ff 1 6 \ee 2 3 \ee 2 5 \ff 2 4 \ff 2 3 \ff 5 6 \ff 3 6 \ff 4 6 \ff 2 7 \ee 3 7 \ee 5 7 \ff 4 7 \ff 3 7 \ff 6 7
\end{pgfonlayer}
\end{scope}
\end{tikzpicture}
$};
\node (b) at (7.5,3) {$\def\xo{.3}
\begin{tikzpicture}[scale=\microscale]
\node [v] (1) at (1,-.8) {$\bu$};
\node [v] (7) at (1,2.8) {$$\bu$$};
\node [v] (2) at (-1,1) {$\bu$};
\node [v] (3) at (0,1) {$x$};
\node [v] (4) at (-\xo,1+\yo) {$\bu$};
\node [v] (5) at (\xo,1-\yo) {$\bu$};
\node [v] (6) at (1,1) {$\bu$};
\begin{pgfonlayer}{background}
\ff 1 2 \ff 1 3 \ff 1 4 \ff 1 5 \ff 1 6 \ee 2 3 \ee 2 5 \ff 2 4 \ff 2 3 \ff 5 6 \ff 3 6 \ff 4 6 \ff 2 7 \ee 3 7 \ee 5 7 \ff 4 7 \ff 3 7 \ff 6 7
\end{pgfonlayer}
\node [v] (11) at (2,1) {$y$};
\node [v] (8) at (2-\xo,1+\yo) {$\bu$};
\node [v] (9) at (2+\xo,1-\yo) {$\bu$};
\node [v] (10) at (3,1) {$\bu$};
\begin{pgfonlayer}{background}
\ff 1 9 \ff 1 {11} \ff 1 8 \ff 6 8 \ee 6 {11} \ee 6 9 \ff 8 {10} \ff {11} {10} \ff 9 {10} \ff {10} 7 \ff 8 7 \ff {10} 1
\end{pgfonlayer}
\ee {11} 7 \ee 9 7
\node [v] (11) at (2,1) {$y$};
\end{tikzpicture}
$};

\node (c) at (5,6) {$\begin{tikzpicture}[scale=\microscale]
\node [v] (1) at (0,0) {$\bu$};
\node [v] (2) at (-1,1) {$\bu$};
\node [v] (3) at (0,1) {$y$};
\node [v] (4) at (-\xo,1+\yo) {$\bu$};
\node [v] (5) at (\xo,1-\yo) {$\bu$};
\node [v] (6) at (1,1) {$\bu$};
\node [v] (7) at (0,2) {$\bu$};
\begin{pgfonlayer}{background}
\ff 1 2 \ff 1 3 \ff 1 4 \ff 1 5 \ff 1 6 \ee 2 3 \ee 2 5 \ff 2 4 \ff 2 3 \ff 5 6 \ff 3 6 \ff 4 6 \ff 2 7 \ee 3 7 \ee 5 7 \ff 4 7 \ff 3 7 \ff 6 7
\end{pgfonlayer}
\begin{scope}[yshift=2cm]
\node [v, white, opacity=0] (1) at (0,0) {7};
\node [v] (2) at (-1,1) {\bu};
\node [v] (3) at (0,1) {$x$};
\node [v] (4) at (-\xo,1+\yo) {\bu};
\node [v] (5) at (\xo,1-\yo) {$\bu$};
\node [v] (6) at (1,1) {\bu};
\node [v] (7) at (0,2) {\bu};
\begin{pgfonlayer}{background}
\ff 1 2 \ff 1 3 \ff 1 4 \ff 1 5 \ff 1 6 \ee 2 3 \ee 2 5 \ff 2 4 \ff 2 3 \ff 5 6 \ff 3 6 \ff 4 6 \ff 2 7 \ee 3 7 \ee 5 7 \ff 4 7 \ff 3 7 \ff 6 7
\end{pgfonlayer}
\end{scope}
\end{tikzpicture}
$};


\node (a) at (15,0) {$\begin{tikzpicture}[scale=\microscale]
\node [v] (1) at (0,0) {$\bu$};
\node [v] (2) at (-1,1) {$\bu$};
\node [v] (3) at (0,1) {$x$};
\node [v] (4) at (-\xo,1+\yo) {$\bu$};
\node [v] (5) at (\xo,1-\yo) {$\bu$};
\node [v] (6) at (1,1) {$\bu$};
\node [v] (7) at (0,2) {$\bu$};
\begin{pgfonlayer}{background}
\ff 1 2 \ff 1 3 \ff 1 4 \ff 1 5 \ff 1 6 \ee 2 3 \ee 2 5 \ff 2 4 \ff 2 3 \ff 5 6 \ff 3 6 \ff 4 6 \ff 2 7 \ee 3 7 \ee 5 7 \ff 4 7 \ff 3 7 \ff 6 7
\end{pgfonlayer}
\begin{scope}[yshift=2cm]
\node [v, white, opacity=0] (1) at (0,0) {7};
\node [v] (2) at (-1,1) {\bu};
\node [v] (3) at (0,1) {$y$};
\node [v] (4) at (-\xo,1+\yo) {\bu};
\node [v] (5) at (\xo,1-\yo) {$\bu$};
\node [v] (6) at (1,1) {\bu};
\node [v] (7) at (0,2) {\bu};
\begin{pgfonlayer}{background}
\ff 1 2 \ff 1 3 \ff 1 4 \ff 1 5 \ff 1 6 \ee 2 3 \ee 2 5 \ff 2 4 \ff 2 3 \ff 5 6 \ff 3 6 \ff 4 6 \ff 2 7 \ee 3 7 \ee 5 7 \ff 4 7 \ff 3 7 \ff 6 7
\end{pgfonlayer}
\end{scope}
\end{tikzpicture}
$};
\node (b) at (12.5,3) {$\def\xo{.3}
\begin{tikzpicture}[scale=\microscale]
\node [v] (1) at (0,-.8) {$\bu$};
\node [v] (2) at (-1-\xo,1+\yo) {$\bu$};
\node [v] (3) at (-\xo,1+\yo) {$x$};
\node [v] (4) at (-2*\xo,1+2*\yo) {$\bu$};
\node [v] (5) at (0,1) {$\bu$};
\node [v] (6) at (1-\xo,1+\yo) {$\bu$};
\node [v] (7) at (0,2.8) {$\bu$};
\begin{pgfonlayer}{background}
\ff 1 2 \ff 1 3 \ff 1 4 \ff 1 5 \ff 1 6 \ee 2 3 \ee 2 5 \ff 2 4 \ff 2 3 \ff 5 6 \ff 3 6 \ff 4 6 \ff 2 7 \ee 3 7 \ee 5 7 \ff 4 7 \ff 3 7 \ff 6 7
\end{pgfonlayer}
\node [v] (11) at (\xo-1,1-\yo) {$\bu$};
\node [v] (8) at (\xo,1-\yo) {$y$};
\node [v] (9) at (2*\xo,1-2*\yo) {$\bu$};
\node [v] (10) at (1+\xo,1-\yo) {$\bu$};
\begin{pgfonlayer}{background}
\ee 5 {11} \ee 5 {10} \ee 8 {10} \ee 9 {10} \ee 8 {11} \ee 9 {11} \ff 1 {11} \ff 1 9 \ff 1 8 \ff 1 {10} \ee {11} 7 \ee 8 7 \ee 9 7 \ee {10} 7
\end{pgfonlayer}
\end{tikzpicture}
$};
\node (c) at (10,6) {$\begin{tikzpicture}[scale=\microscale]
\node [v] (1) at (0,0) {$\bu$};
\node [v] (2) at (-1,1) {\bu};
\node [v] (3) at (0,1) {$y$};
\node [v] (4) at (-\xo,1+\yo) {$\bu$};
\node [v] (5) at (\xo,1-\yo) {$\bu$};
\node [v] (6) at (1,1) {$\bu$};
\node [v] (7) at (0,2) {$\bu$};
\begin{pgfonlayer}{background}
\ff 1 2 \ff 1 3 \ff 1 4 \ff 1 5 \ff 1 6 \ee 2 3 \ee 2 5 \ff 2 4 \ff 2 3 \ff 5 6 \ff 3 6 \ff 4 6 \ff 2 7 \ee 3 7 \ee 5 7 \ff 4 7 \ff 3 7 \ff 6 7
\end{pgfonlayer}
\begin{scope}[yshift=2cm]
\node [v, white, opacity=0] (1) at (0,0) {7};
\node [v] (2) at (-1,1) {$\bu$};
\node [v] (3) at (0,1) {$x$};
\node [v] (4) at (-\xo,1+\yo) {$\bu$};
\node [v] (5) at (\xo,1-\yo) {$\bu$};
\node [v] (6) at (1,1) {$\bu$};
\node [v] (7) at (0,2) {$\bu$};
\begin{pgfonlayer}{background}
\ff 1 2 \ff 1 3 \ff 1 4 \ff 1 5 \ff 1 6 \ee 2 3 \ee 2 5 \ff 2 4 \ff 2 3 \ff 5 6 \ff 3 6 \ff 4 6 \ff 2 7 \ee 3 7 \ee 5 7 \ff 4 7 \ff 3 7 \ff 6 7
\end{pgfonlayer}
\end{scope}
\end{tikzpicture}
$};

\draw [->, thick] (4.3,3) to +(.8,0);
\draw [->, thick] (10.8,3) to +(-.8,0);
\draw [->, thick] (6.5,0) to +(2,0);
\draw [->, thick] (13.5,0) to +(-2,0);
\draw [->, thick] (1.5,6) to +(2,0);
\draw [->, thick] (8.5,6) to +(-2,0);
\draw [->, thick] (4,1.24) to +(-.5,.6);
\draw [->, thick] (9,1.24) to +(-.5,.6);
\draw [->, thick] (14,1.24) to +(-.5,.6);
\draw [<-, thick] (1.6,4.16) to +(-.5,.6);
\draw [<-, thick] (6.7,4.16) to +(-.5,.6);
\draw [<-, thick] (11.7,4.16) to +(-.5,.6);



\end{tikzpicture}
$
\caption{\label{fig:syllepsis}The zigzag structure of the syllepsis as a 5-diagram.}
\end{figure*}

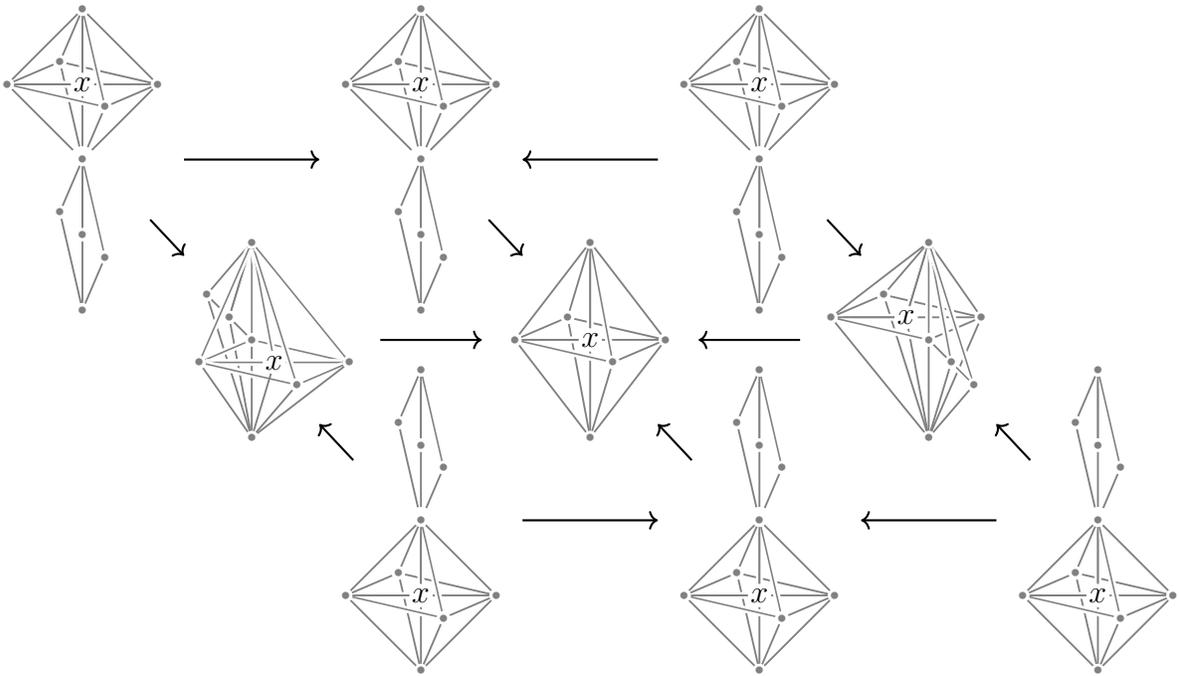
\begin{figure*}[h!]
$
\def\microscale{1}
\begin{tikzpicture}[yscale=.8, xscale=.9]


\node [scale=\microscale] (a) at (5,0) {$\begin{tikzpicture}
\node [v] (1) at (0,0) {$\bu$};
\node [v] (2) at (-1,1) {$\bu$};
\node [v] (3) at (0,1) {$x$};
\node [v] (4) at (-\xo,1+\yo) {$\bu$};
\node [v] (5) at (\xo,1-\yo) {$\bu$};
\node [v] (6) at (1,1) {$\bu$};
\node [v] (7) at (0,2) {$\bu$};
\begin{pgfonlayer}{background}
\ff 1 2 \ff 1 3 \ff 1 4 \ff 1 5 \ff 1 6 \ee 2 3 \ee 2 5 \ff 2 4 \ff 2 3 \ff 5 6 \ff 3 6 \ff 4 6 \ff 2 7 \ee 3 7 \ee 5 7 \ff 4 7 \ff 3 7 \ff 6 7
\end{pgfonlayer}
\begin{scope}[yshift=2cm]
\node [v, white, opacity=0] (1) at (0,0) {7};
\node [v] (3) at (0,1) {$\bu$};
\node [v] (4) at (-\xo,1+\yo) {\bu};
\node [v] (5) at (\xo,1-\yo) {$\bu$};
\node [v] (7) at (0,2) {\bu};
\begin{pgfonlayer}{background}
\ff 1 3 \ff 1 4 \ff 1 5 \ee 3 7 \ee 5 7 \ff 4 7 \ff 3 7
\end{pgfonlayer}
\end{scope}
\end{tikzpicture}
$};
\node [scale=\microscale] (b) at (2.5,3) {$\def\xo{.3}
\begin{tikzpicture}[xscale=1]
\node [v] (1) at (0,-.3) {$\bu$};
\node [v] (3) at (-\xo,1+\yo) {$\bu$};
\node [v] (4) at (-2*\xo,1+2*\yo) {$\bu$};
\node [v] (5) at (0,1) {$\bu$};
\node [v] (7) at (0,2.3) {$\bu$};
\begin{pgfonlayer}{background}
\ff 3 4 \ff 5 3 \ff 1 3 \ff 1 4 \ff 1 5 \ee 3 7 \ee 5 7 \ff 4 7 \ff 3 7
\end{pgfonlayer}
\node [v] (11) at (\xo-1,1-\yo) {$\bu$};
\node [v] (8) at (\xo,1-\yo) {$x$};
\node [v] (9) at (2*\xo,1-2*\yo) {$\bu$};
\node [v] (10) at (1+\xo,1-\yo) {$\bu$};
\begin{pgfonlayer}{background}
\ee 5 {11} \ee 5 {10} \ee 8 {10} \ee 9 {10} \ee 8 {11} \ee 9 {11} \ff 1 {11} \ff 1 9 \ff 1 8 \ff 1 {10} \ee {11} 7 \ee 8 7 \ee 9 7 \ee {10} 7
\end{pgfonlayer}
\node [v] (10) at (-1-\xo,1-\yo) {\phantom{$\bu$}};
\end{tikzpicture}
$};
\node [scale=\microscale] (c) at (0,6) {$\begin{tikzpicture}
\node [v] (1) at (0,0) {$\bu$};
\node [v] (3) at (0,1) {$\bu$};
\node [v] (4) at (-\xo,1+\yo) {$\bu$};
\node [v] (5) at (\xo,1-\yo) {$\bu$};
\node [v] (7) at (0,2) {$\bu$};
\begin{pgfonlayer}{background}
\ff 1 3 \ff 1 4 \ff 1 5 \ee 3 7 \ee 5 7 \ff 4 7 \ff 3 7
\end{pgfonlayer}
\begin{scope}[yshift=2cm]
\node [v, white, opacity=0] (1) at (0,0) {7};
\node [v] (2) at (-1,1) {$\bu$};
\node [v] (3) at (0,1) {$x$};
\node [v] (4) at (-\xo,1+\yo) {$\bu$};
\node [v] (5) at (\xo,1-\yo) {$\bu$};
\node [v] (6) at (1,1) {$\bu$};
\node [v] (7) at (0,2) {$\bu$};
\begin{pgfonlayer}{background}
\ff 1 2 \ff 1 3 \ff 1 4 \ff 1 5 \ff 1 6 \ee 2 3 \ee 2 5 \ff 2 4 \ff 2 3 \ff 5 6 \ff 3 6 \ff 4 6 \ff 2 7 \ee 3 7 \ee 5 7 \ff 4 7 \ff 3 7 \ff 6 7
\end{pgfonlayer}
\end{scope}
\end{tikzpicture}
$};


\node [scale=\microscale] (a) at (10,0) {$\begin{tikzpicture}
\node [v] (1) at (0,0) {$\bu$};
\node [v] (2) at (-1,1) {$\bu$};
\node [v] (3) at (0,1) {$x$};
\node [v] (4) at (-\xo,1+\yo) {$\bu$};
\node [v] (5) at (\xo,1-\yo) {$\bu$};
\node [v] (6) at (1,1) {$\bu$};
\node [v] (7) at (0,2) {$\bu$};
\begin{pgfonlayer}{background}
\ff 1 2 \ff 1 3 \ff 1 4 \ff 1 5 \ff 1 6 \ee 2 3 \ee 2 5 \ff 2 4 \ff 2 3 \ff 5 6 \ff 3 6 \ff 4 6 \ff 2 7 \ee 3 7 \ee 5 7 \ff 4 7 \ff 3 7 \ff 6 7
\end{pgfonlayer}
\begin{scope}[yshift=2cm]
\node [v, white, opacity=0] (1) at (0,0) {7};
\node [v] (3) at (0,1) {$\bu$};
\node [v] (4) at (-\xo,1+\yo) {\bu};
\node [v] (5) at (\xo,1-\yo) {$\bu$};
\node [v] (7) at (0,2) {\bu};
\begin{pgfonlayer}{background}
\ff 1 3 \ff 1 4 \ff 1 5 \ee 3 7 \ee 5 7 \ff 4 7 \ff 3 7
\end{pgfonlayer}
\end{scope}
\end{tikzpicture}
$};
\node [scale=\microscale] (b) at (7.5,3) {$\def\xo{.3}
\begin{tikzpicture}[xscale=1]
\node [v] (1) at (0,-.3) {$\bu$};
\node [v] (7) at (0,2.3) {$$\bu$$};
\node [v] (2) at (-1,1) {$\bu$};
\node [v] (3) at (0,1) {$x$};
\node [v] (4) at (-\xo,1+\yo) {$\bu$};
\node [v] (5) at (\xo,1-\yo) {$\bu$};
\node [v] (6) at (1,1) {$\bu$};
\begin{pgfonlayer}{background}
\ff 1 2 \ff 1 3 \ff 1 4 \ff 1 5 \ff 1 6 \ee 2 3 \ee 2 5 \ff 2 4 \ff 2 3 \ff 5 6 \ff 3 6 \ff 4 6 \ff 2 7 \ee 3 7 \ee 5 7 \ff 4 7 \ff 3 7 \ff 6 7
\end{pgfonlayer}
\begin{pgfonlayer}{background}
\end{pgfonlayer}
\end{tikzpicture}
$};

\node [scale=\microscale] (c) at (5,6) {$\begin{tikzpicture}
\node [v] (1) at (0,0) {$\bu$};
\node [v] (3) at (0,1) {$\bu$};
\node [v] (4) at (-\xo,1+\yo) {$\bu$};
\node [v] (5) at (\xo,1-\yo) {$\bu$};
\node [v] (7) at (0,2) {$\bu$};
\begin{pgfonlayer}{background}
\ff 1 3 \ff 1 4 \ff 1 5 \ee 3 7 \ee 5 7 \ff 4 7 \ff 3 7
\end{pgfonlayer}
\begin{scope}[yshift=2cm]
\node [v, white, opacity=0] (1) at (0,0) {7};
\node [v] (2) at (-1,1) {$\bu$};
\node [v] (3) at (0,1) {$x$};
\node [v] (4) at (-\xo,1+\yo) {$\bu$};
\node [v] (5) at (\xo,1-\yo) {$\bu$};
\node [v] (6) at (1,1) {$\bu$};
\node [v] (7) at (0,2) {$\bu$};
\begin{pgfonlayer}{background}
\ff 1 2 \ff 1 3 \ff 1 4 \ff 1 5 \ff 1 6 \ee 2 3 \ee 2 5 \ff 2 4 \ff 2 3 \ff 5 6 \ff 3 6 \ff 4 6 \ff 2 7 \ee 3 7 \ee 5 7 \ff 4 7 \ff 3 7 \ff 6 7
\end{pgfonlayer}
\end{scope}
\end{tikzpicture}
$};


\node [scale=\microscale] (a) at (15,0) {$\begin{tikzpicture}
\node [v] (1) at (0,0) {$\bu$};
\node [v] (2) at (-1,1) {$\bu$};
\node [v] (3) at (0,1) {$x$};
\node [v] (4) at (-\xo,1+\yo) {$\bu$};
\node [v] (5) at (\xo,1-\yo) {$\bu$};
\node [v] (6) at (1,1) {$\bu$};
\node [v] (7) at (0,2) {$\bu$};
\begin{pgfonlayer}{background}
\ff 1 2 \ff 1 3 \ff 1 4 \ff 1 5 \ff 1 6 \ee 2 3 \ee 2 5 \ff 2 4 \ff 2 3 \ff 5 6 \ff 3 6 \ff 4 6 \ff 2 7 \ee 3 7 \ee 5 7 \ff 4 7 \ff 3 7 \ff 6 7
\end{pgfonlayer}
\begin{scope}[yshift=2cm]
\node [v, white, opacity=0] (1) at (0,0) {7};
\node [v] (3) at (0,1) {$\bu$};
\node [v] (4) at (-\xo,1+\yo) {\bu};
\node [v] (5) at (\xo,1-\yo) {$\bu$};
\node [v] (7) at (0,2) {\bu};
\begin{pgfonlayer}{background}
\ff 1 3 \ff 1 4 \ff 1 5 \ee 3 7 \ee 5 7 \ff 4 7 \ff 3 7
\end{pgfonlayer}
\end{scope}
\end{tikzpicture}
$};
\node [scale=\microscale] (b) at (12.5,3) {$\def\xo{.3}
\begin{tikzpicture}[xscale=1]
\node [v] (1) at (0,-.3) {$\bu$};
\node [v] (2) at (-1-\xo,1+\yo) {$\bu$};
\node [v] (3) at (-\xo,1+\yo) {$x$};
\node [v] (4) at (-2*\xo,1+2*\yo) {$\bu$};
\node [v] (5) at (0,1) {$\bu$};
\node [v] (6) at (1-\xo,1+\yo) {$\bu$};
\node [v] (7) at (0,2.3) {$\bu$};
\begin{pgfonlayer}{background}
\ff 1 2 \ff 1 3 \ff 1 4 \ff 1 5 \ff 1 6 \ee 2 3 \ee 2 5 \ff 2 4 \ff 2 3 \ff 5 6 \ff 3 6 \ff 4 6 \ff 2 7 \ee 3 7 \ee 5 7 \ff 4 7 \ff 3 7 \ff 6 7
\end{pgfonlayer}
\node [v] (8) at (\xo,1-\yo) {$\bu$};
\node [v] (9) at (2*\xo,1-2*\yo) {$\bu$};
\begin{pgfonlayer}{background}
\ff 1 9 \ff 1 8 \ee 8 7 \ee 9 7 \ff 5 8 \ee 8 9
\end{pgfonlayer}
\node [v] (2) at (1+\xo,1+\yo) {\phantom{$\bu$}};
\end{tikzpicture}
$};

\node [scale=\microscale] (c) at (10,6) {$\begin{tikzpicture}
\node [v] (1) at (0,0) {$\bu$};
\node [v] (3) at (0,1) {$\bu$};
\node [v] (4) at (-\xo,1+\yo) {$\bu$};
\node [v] (5) at (\xo,1-\yo) {$\bu$};
\node [v] (7) at (0,2) {$\bu$};
\begin{pgfonlayer}{background}
\ff 1 3 \ff 1 4 \ff 1 5 \ee 3 7 \ee 5 7 \ff 4 7 \ff 3 7
\end{pgfonlayer}
\begin{scope}[yshift=2cm]
\node [v, white, opacity=0] (1) at (0,0) {7};
\node [v] (2) at (-1,1) {$\bu$};
\node [v] (3) at (0,1) {$x$};
\node [v] (4) at (-\xo,1+\yo) {$\bu$};
\node [v] (5) at (\xo,1-\yo) {$\bu$};
\node [v] (6) at (1,1) {$\bu$};
\node [v] (7) at (0,2) {$\bu$};
\begin{pgfonlayer}{background}
\ff 1 2 \ff 1 3 \ff 1 4 \ff 1 5 \ff 1 6 \ee 2 3 \ee 2 5 \ff 2 4 \ff 2 3 \ff 5 6 \ff 3 6 \ff 4 6 \ff 2 7 \ee 3 7 \ee 5 7 \ff 4 7 \ff 3 7 \ff 6 7
\end{pgfonlayer}
\end{scope}
\end{tikzpicture}
$};

\draw [->, thick] (4.4,3) to +(1.5,0);
\draw [->, thick] (10.6,3) to +(-1.5,0);
\draw [->, thick] (6.5,0) to +(2,0);
\draw [->, thick] (13.5,0) to +(-2,0);
\draw [->, thick] (1.5,6) to +(2,0);
\draw [->, thick] (8.5,6) to +(-2,0);
\draw [->, thick] (4,1) to +(-.5,.6);
\draw [->, thick] (9,1) to +(-.5,.6);
\draw [->, thick] (14,1) to +(-.5,.6);
\draw [<-, thick] (1.5,4.4) to +(-.5,.6);
\draw [<-, thick] (6.5,4.4) to +(-.5,.6);
\draw [<-, thick] (11.5,4.4) to +(-.5,.6);

\end{tikzpicture}
$
\caption{\label{fig:syllepsis_x}The singular piece containing the generator $x$ in the zigzag structure of the syllepsis.}
\end{figure*}

\clearpage

\bibliographystyle{plainurl}
\bibliography{arxiv}

\end{document}